\documentclass[fleqn]{2020SCGE}
\usepackage{color}
\usepackage{natbib}
\usepackage{gensymb}
\usepackage{graphicx}
\usepackage{txfonts}
\usepackage{aalongtable}
\usepackage{url}
\usepackage{hyperref}%PDF
\def\degr{\hbox{$^\circ$}}
\def\arcmin{\hbox{$^\prime$}}
\def\arcsec{\hbox{$^{\prime\prime}$}}
\def\fdg{\hbox{$.\!\!^\circ$}}
\def\farcm{\hbox{$.\mkern-4mu^\prime$}}
\DeclareRobustCommand{\ion}[2]{\textup{#1\,\textsc{\lowercase{#2}}}}

\begin{document}
%\linenumbers

\ensubject{subject}

%%%%%%%%%%%%%%%%%%%%%%%%%%%%%%%%%%%%%%%%%%%%%%%%%%%%%%%
%%% Authors do not modify the information below

%Letter to the Editor
\ArticleType{Article}
\SpecialTopic{Special Topic: Peering into the Milky Way by FAST}
\Year{2022}
\Month{December}
\Vol{65}
\No{12}
\DOI{10.1007/s11433-022-2031-7}
\ArtNo{129705}
\ReceiveDate{July 14, 2022}
\AcceptDate{November 9, 2022}
\OnlineDate{November 24, 2022}
%%%%%%%%%%%%%%%%%%%%%%%%%%%%%%%%%%%%%%%%%%%%%%%%%%%%%%%

\title{Peering into the Milky Way by FAST: 
\\ IV. Identification of two new Galactic supernova remnants G203.1+6.6 and G206.7+5.9}{Peering into the Milky Way by FAST:  IV. Identification of two new Galactic supernova remnants G203.1+6.6 and G206.7+5.9}

%%% Corresponding author: 
%%%   \author[number]{Full name}{{email@xxx.com}}
%%% General author: 
%%%   \author[number]{Full name}{}
%\author[]{{\color{red}{(The author names, affiliations and ranking will be finalized after the Fermi collaboration review.)}}}{}

\author[1,5]{XuYang~Gao}{{xygao@nao.cas.cn}}
\author[2]{Wolfgang~Reich}{}
\author[3]{XiaoHui~Sun}{{xhsun@ynu.edu.cn}}
\author[4]{He~Zhao}{}
\author[1]{Tao~Hong}{}
\author[1]{ZhongSheng~Yuan}{}
\author[2]{\\Patricia Reich}{}
\author[1,5]{JinLin~Han}{{hjl@nao.cas.cn}}

\footnotetext[1]{* Corresponding Authors (XuYang~Gao, email:xygao@nao.cas.cn, XiaoHui~Sun, email:xhsun@ynu.edu.cn, JinLin~Han, email:jlhan@nao.cas.cn}

%%% Add a version number to the draft (remove before submission)
%\footnote{Version 8, Update date: 2020/03/28}

%%% Author information for page head.
%\AuthorMark{Xuyang~Gao}
\AuthorMark{X. Y. Gao}
%%% Authors for citation. 
\AuthorCitation{X. Y.~Gao, W. Reich, X. H.~Sun, H. Zhao, T. Hong, Z. S.~Yuan, P. Reich, and J. L.~Han}

%%% Address.
%%% \address[number]{Address, City {\rm Postcode}, Country}
\address[{\rm1}]{National Astronomical Observatories, Chinese Academy
  of Sciences, Beijing 100101, PR China}
\address[{\rm2}]{Max-Planck-Institut f{\"u}r Radioastronomie, 53121 Bonn, Germany}
\address[{\rm3}]{School of Physics and Astronomy, Yunnan University,
  Kunming 650500, PR China}
\address[{\rm4}]{Purple Mountain Observatory, Chinese Academy of
  Sciences, Nanjing 210023, PR China}
\address[{\rm5}]{School of Astronomy, University of Chinese Academy of
  Sciences, Beijing 100049, PR China}

%%% Abstract.
\abstract{A $5\degr\times7\degr$ sky area containing two large radio
  structures of G203.1+6.6 and G206.7+5.9 with a size of about
  2$\fdg$5 and 3$\fdg$5 respectively is scanned by using the L-band
  19-beam receiver of the Five-hundred-meter Aperture Spherical radio
  Telescope (FAST). The FAST L-band receiver covers a frequency range
  of 1.0-1.5~GHz. Commissioning of the receiving system,
  including the measurements of the half-power beam width, gain, and
  main-beam efficiency is made by observing the calibrators. The
  multi-channel spectroscopy backend mounted to the receiver allows an
  in-band spectral-index determination. The brightness-temperature
  spectral indices of both objects are measured to be $\beta \sim
  -2.6$ to $-2.7$. Polarized emission is detected from the archival
  Effelsberg $\lambda$11\ cm data for all the shell structures of
  G203.1+6.6 and G206.7+5.9. These results clearly indicate a
  non-thermal synchrotron emitting nature, confirming that G203.1+6.6
  and G206.7+5.9 are large shell-type supernova remnants (SNRs). Based
  on morphological correlation between the radio continuum emission of
  G206.7+5.9 and the \ion{H}{I} structures, the kinematic distance to
  this new SNR is estimated to be about 440~pc, placing it in the
  Local Arm.
}%

%%% Keywords.
%\keywords{ISM: supernova remnants --- ISM: magnetic fields ---
%  polarization --- techniques: Polarimetry}

\maketitle
%%%\tableofcontents%

\begin{multicols}{2}

%%%%%%%%%%%%%%%%%%%%%%%%%%%%%%%%%%%%%%%%%%%%%%%%%%%%%%%%%%%%
%% Text of article.
%%%%%%%%%%%%%%%%%%%%%%%%%%%%%%%%%%%%%%%%%%%%%%%%%%%%%%%%%%%%

\section{Introduction}           
\label{sect:intro}
The Five-hundred-meter Aperture Spherical radio Telescope
\citep[FAST,][]{nan08,nlj+11}, mounted with the L-band 19-beam
receiver, is an extremely sensitive radio telescope to observe pulsars
\citep{Han21}, the spectral line of atomic hydrogen \citep{Hong22},
and the radio recombination lines of ionized gas \citep{Hou22}. We
work on a series of papers which are dedicated to investigations of
the Galactic interstellar medium (ISM) by FAST. The first two papers
\citep{Hong22, Hou22} are based on the spectral-line data
simultaneously recorded by the Galactic Plane Pulsar Snapshot (GPPS)
survey \citep{Han21} of the FAST accessible sky with $|b| \leq
10\degr$. \citet[][Paper I]{Hong22} focused on the \ion{H}{I}
structures from the high-resolution and high-sensitivity piggyback
\ion{H}{I}-line observations of the FAST GPPS survey, while
\citet[][Paper II]{Hou22} studied the radio recombination lines of
ionized gas. In the third paper, \citet{Xu22} revealed the
interstellar magnetic fields by measuring the Faraday effect of a
large number of weak pulsars. The work presented here is the fourth in
the series. We scan a $5\degr\times7\degr$ sky area to get the radio
continuum image by using the L-band 19-beam receiver of FAST, and
identify two new large supernova remnants (SNRs) G203.1+6.6 and
G206.7+5.9.

SNRs are prominent radio sources that play an important role in the
Galactic ecological system by processing the interstellar medium. Up
to now, about 300 SNRs have been discovered in the Galaxy
\citep{Green19}, most of which are identified based on radio
observations \citep[e.g.][]{Dubner15}. Despite occasional reports on
some individual or a small group of new SNRs \citep[e.g][]{Gao11y,
  Foster13, Sabin13, Gao14, Kothes14, Gerbrandt14}, numerous Galactic
SNRs are still waiting to be discovered.

In the last few years, sensitive and high angular resolution radio
observations utilizing synthesis arrays bring a boost. With an angular
resolution down to $\sim$20$\arcsec$, The \ion{HI}/OH/Recombination
line survey of the inner Milky Way (THOR) project \citep{Anderson17}
surveyed a narrow section of the inner Galactic plane of $14\fdg5 < l
< 67\fdg4$, and $|b| < 1\fdg25$ by using the Very Large Array (VLA),
and identified 76 new SNR candidates. Another project conducted with
the VLA but working at the C-band \citep[4-8~GHz, GLOSTAR
  project,][]{Dokara21} discovered another 80 new SNR candidates
within the Galactic plane area of $358\degr < l < 60\degr$ and $|b| <
1\degr$. With the Murchison Widefield Array operating in the frequency
range of 72-231~MHz, several tens of new SNRs have been
identified in the Galactic plane area of $345\degr < l < 60\degr,
180\degr < l < 240\degr$, and $|b| \leqslant 10\degr$
\citep{Hurley-Walker19a, Hurley-Walker19b}. Recently,
\citet{Heywood22} surveyed a 6.5-$\rm deg^2$ sky area at the Galactic
center by using MeerKAT with an extremely high angular resolution of
4$\arcsec$, a few new SNR candidates and many radio arcs have been
revealed.

Besides these SNRs/SNR candidates discovered in the narrow stripes of
the Galactic plane, some SNRs have been uncovered at higher Galactic
latitude, e.g. G159.6+7.3 \citep{Fesen10}, G70.0$-$21.5 \citep{Fesen15},
G181.1+9.5 \citep{Kothes17}, G107.0+9.0 \citep{Fesen20}, and
G17.8+16.7 \citep{Araya22}. They are out of the sky coverage of major
Galactic plane surveys, which are often limited up to $|b| = 5\degr$,
e.g. the Effelsberg $\lambda$11\ cm survey \citep{Fuerst90,
  Reich9011}, the Sino-German $\lambda$6\ cm survey \citep{Sun07,
  Gao10, Sun11a, Xiao11}, and the 1.4-GHz southern Galactic plane
survey \citep{Haverkorn06}. Some Galactic plane surveys have extended
to high latitudes in selected areas. For example, the Canadian
Galactic Plane Survey \citep{Taylor03, Landecker10} extended to $b =
+17\fdg5$ in the Galactic longitude range of $l = 101\degr$-$116\degr$. 
The 2.4-GHz southern Galactic plane survey extended in some
of its coverage to $b = +7\degr$ and $b = -8\degr$
\citep{Duncan97}. The Effelsberg Medium Latitude Survey
\citep[EMLS,][]{Uyaniker98, Reich04} conducted with the Effelsberg
100-m radio telescope aims to cover a Galactic latitude range of $|b|
< 20\degr$ in its accessible sky. The first published section released
the data in four fields, covering 1\,100~$\rm deg^2$ in total
\citep{Uyaniker99}.
From the unpublished data of EMLS in the Galactic anti-center area,
\citet{Reich02} presented two SNR candidates, G203.1+6.6 and
G206.7+5.9, both of which show limb-brightened shell
structures. However, the narrow-band total-intensity observations
without polarization measurements did not allow to identify the nature
of these two objects.

FAST is the largest single-dish telescope in the world. Inspired in
utilizing FAST for radio-continuum study, we develop the FAST
calibration and data-processing procedures, and scan a
$5\degr\times7\degr$ sky area which contains G203.1+6.6 and G206.7+5.9
with the FAST L-band 19-beam receiver. The paper is organized as
follows. In Sect.~\ref{sect:ObsD}, we introduce the FAST observations
and data reduction. We present the observational results and distance
estimate in Sect.~\ref{sect:results}. The conclusions are given in
Sect.~\ref{sect:conclusion}.

\section{Observations and data reduction}
\label{sect:ObsD}

\begin{figure}[H]
\centering
\includegraphics[width=0.475\textwidth]{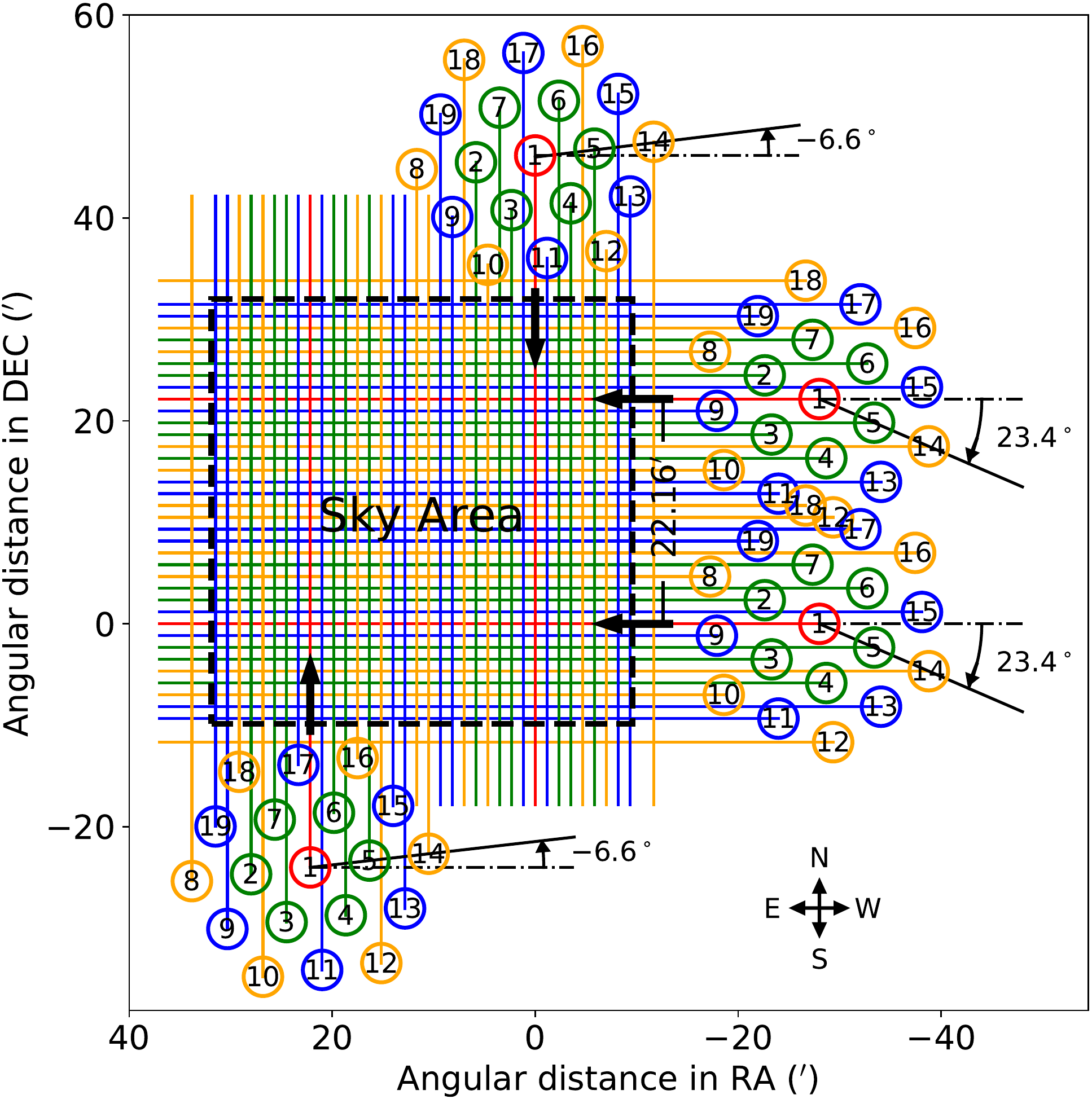}
\caption{Schematic plot of FAST drift-scan and ``MultiBeam OTF''
  observation modes for continuum imaging in the sky plane. Beam
  numbers are indicated in each beam. The hexagon 19-beam system is
  rotated by $+23\fdg4$ (clockwise) in the drift-scan mode and
  $-6\fdg6$ (counter-clockwise) in the ``MultiBeam OTF'' mode, and
  shifted by 22$\farcm$16 each time to scan the target area. The beams
  with the same color (green, blue and orange) have the same angular
  distance to the central beam (red). The colored lines illustrate the
  tracks of each beam with a separation of 1$\farcm$17.}
\label{mode}
\end{figure}

The FAST observations of G203.1+6.6 and G206.7+5.9 were conducted with
the L-band 19-beam receiver under two observation modes. The
drift-scan mode with the 19-beam receivers rotated by 23$\fdg$4 was
used for scans along the RA direction during the FAST commissioning
phase in late 2019 and 2020 (FAST PID: 0355), while the multi-beam
``On-The-Fly'' (MultiBeam OTF) mode with a rotation angle of
$-$6$\fdg$6 was adopted in 2021 for scans along the DEC direction
during the FAST shared-risk phase (FAST PID: PT2020-0118). Considering
the FAST L-band half-power beam-width (HPBW) of about 3$\arcmin$ at
the higher frequency end of the band \citep{Jiang20}, both modes
provide a full Nyquist sampling with a scan separation of about
1$\farcm$17. A sketch map of the two observation modes is shown in
Fig.~\ref{mode}.

To establish the scale between the antenna temperature and source
flux-density, 3C~138 as the main calibration source 

{\flushleft was observed regularly using the FAST ``Multi-beam 
Calibration'' mode, in which each of the 19 beams tracked the 
calibrator in turn for one minute. To monitor and correct for 
the system-temperature drift, all observations utilized a 
noise tube with a periodically-injected reference signal of about 
1.1~K for the target source and about 12.5~K for the calibrator. 
The scans of the targeted area and the observations toward the 
calibrators took about 20 hours in total.}

\subsection{System characteristics of FAST observations }

The system performance must be verified before observations. The
half-power beam width characterizes the main beam of the telescope
through which majority of the radio emission power is received. The
gain represents the conversion factor from the flux-density scale in
Jy into antenna-temperature scale in K $T_{A}$. Both HPBW and gain are
then used to derive the main-beam efficiency, which further helps
convert the map unit from antenna temperature $T_{A}$ into brightness
temperature $T_{B}$. We discuss these basic system performance
parameters in the following.

\subsubsection{Half-power beam width}

\begin{figure}[H]
\centering
\includegraphics[width=0.475\textwidth]{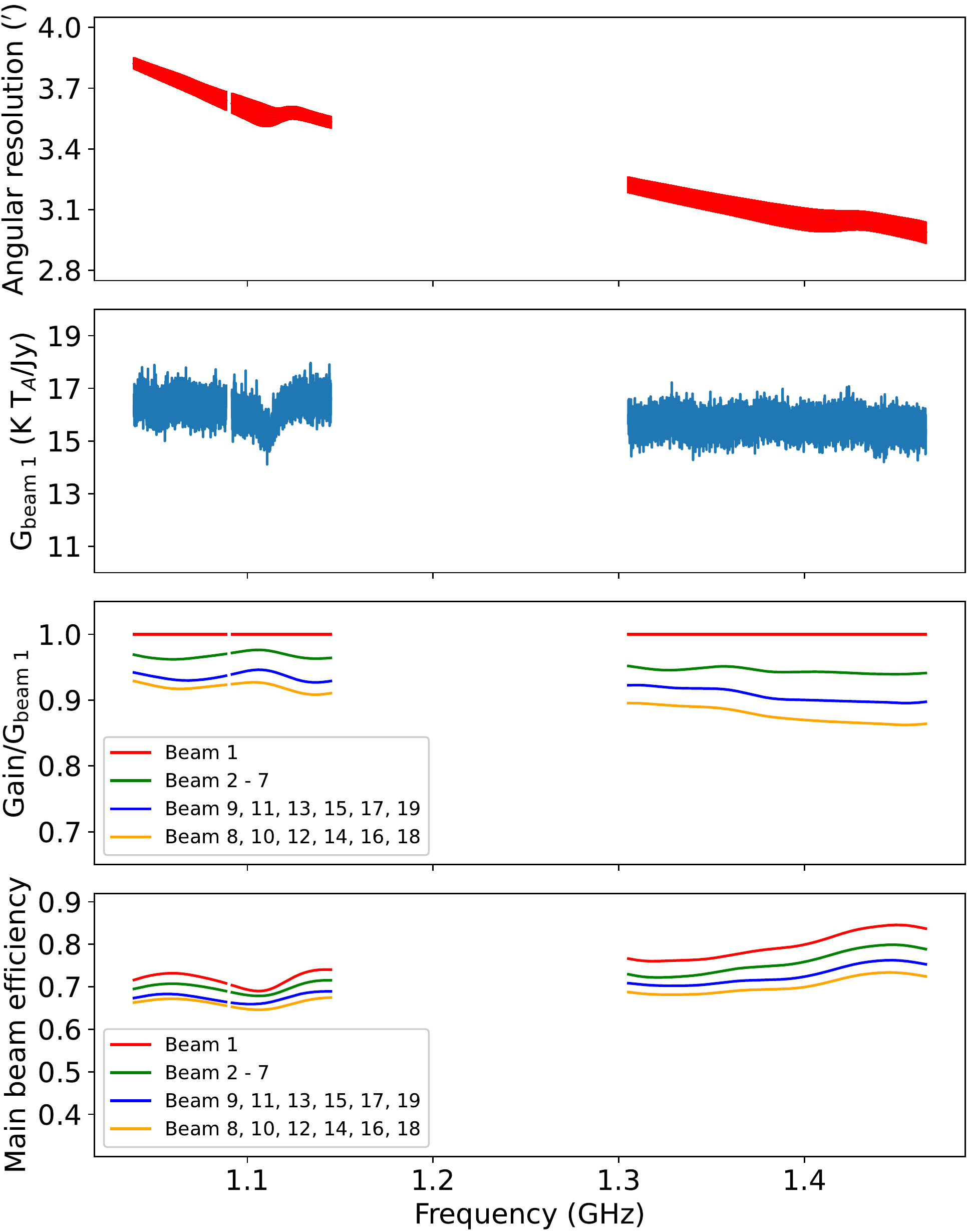}
\caption{The FAST characteristics varying with frequencies. {\it Top
    panel}: FAST HPBW versus observing frequency. The values are
  averaged for all the 19 beams at a given observing frequency. The
  thickness of the red line indicates the standard deviation of the
  values. {\it Middle panels:} Gain of the central beam and the
  normalized gain of the other beams grouped by different angular
  distances from the central beam. The color scheme is the same as in
  Fig.~\ref{mode} (see Fig.~\ref{mode} for detailed beam
  configuration). {\it Bottom panel:} Main-beam efficiency of the 19
  FAST beams, which are grouped following the same scheme as shown in
  the middle panel.}
\label{performance}
\end{figure}

The 19 FAST feeds are hexagonally mounted (see Fig.~\ref{mode} for
reference). Rotating the hexagon to a certain position angle will
align some feeds along the RA direction at the same DEC. Then a drift
scan toward a calibration source can be made to assess the HPBW from
Gaussian fitting, e.g. 0$\degr$ rotation allows the beams No.~8, 2, 1,
5, and 14 to drift across a calibration source, while a +30$\degr$
rotation will align the beams No.~9, 1, and 15. Such measurements were
applied toward several calibrators, i.e.  3C~48, 3C~84, 3C~286, and
3C~380. The HPBW values obtained for the same feed were then
averaged. We finally averaged the HPBWs of all the 19 beams according
to the observing frequencies and show the result in
Fig.~\ref{performance}. At the same observing frequency, the standard
deviation of the HPBW between different beams is less than
2.5\%. Within the usable band, the FAST HPBW was found to be
$\sim3\farcm8$ at the lower-frequency end at 1.039~GHz and about
$3\farcm0$ 

{\flushleft at the higher-frequency end at 1.465~GHz. These results are
compatible with the values found by \citet{Sun21}, but slightly higher
than the values of \citet{Jiang20}. Another feature of the FAST HPBW
is the inversion seen at about 1.11~GHz and 1.41~GHz. This was also
noticed by \citet{Sun21} when estimating FAST HPBW through 2-D
Gaussian fits to the calibrator in their work.}

By measuring the time delay of the fitted Gaussian peaks, we can
estimate the angular separation between the FAST adjacent beams. An
average spacing is 5$\farcm$87, which can be used to determine the
coordinates of the FAST offset beams (Beam No.2 to No.19).

\subsubsection{Gain}
The gain of FAST, defined as $G(\nu) = T(\nu)/S(\nu)$, is essential
for the antenna temperature-flux density conversion (unit:
K\ $T_{A}$/Jy).  It is also of great importance to adjust the
different gain levels of different beams. We estimated $G_{\nu}$ of
FAST based on the flux density $S_{\nu}$ of 3C~138 given in
\citet{Perley17}, in which $S_{\nu}$ of 3C~138 can be obtained from
$log(S_{\nu}) = 1.088 - 0.4981~log(\nu) - 0.1552~[log(\nu)]^2 -
0.0102~[log(\nu)]^3 + 0.0223~[log(\nu)]^4$, where $\nu$ is in GHz, and
$S$ is in Jy. For simplicity in display, we divided the
hexagon-configured 19 FAST beams into four groups: the central beam
No.1 and three outer rings according to their angular distance to Beam
1, i.e. the first ring: Beam Nos. 2-7 with a distance of 5$\farcm$87,
the second ring: Beam Nos. 9, 11, 13, 15, 17, and 19, and the outer-most
ring: Beam Nos. 8, 10, 12, 14, 16, and 18, with an angular distance of
10$\farcm$17 and 11$\farcm$74, respectively (see Fig.~\ref{mode}). The
gains of different beams in the same ring were found nearly identical,
with a standard deviation ($1\sigma$) of about 2\%. We averaged the
gains of beams in the same group and normalized them with respect to
that of the central beam. The result is shown in the middle panels of
Fig.~\ref{performance}. It is evident that the gain decreases with
increasing distance to the center. On average, the gains in the first,
second, and the outer-most rings are about 97\%, 94\%, and 92\% of
that of the central beam at the lower-frequency band, and 95\%, 91\%,
and 88\% at the higher-frequency band. For each single beam, the gain
decreases with the observing frequency.

\begin{table*}[t!]
%\centering
\caption{Parameters of six sources used for verification of the FAST
  data processing. The coordinates (RA, DEC/J2000) were extracted
  from the NVSS source catalog and from the FAST image through 2-D
  Gaussian fitting. The NVSS 1.4-GHz (mean of 1364.9~MHz \&
  1435.1~MHz) and the average value of the FAST 1.391-GHz and
  1.443-GHz flux densities are listed and compared in the 6th and 7th
  rows. The spectral indices determined by fitting the flux-density data
  taken from SPECFIND and the FAST five sub-bands are shown in the 8th
  and 9th rows.\newline
  \label{tab:tab1}}
\begin{tabular}{ccccccc}
  \hline\hline
Source  & \multicolumn{1}{c}{4C~11.24} & \multicolumn{1}{c}{4C~10.21} & \multicolumn{1}{c}{ICRF J065917.9+081330} & \multicolumn{1}{c}{4C~08.24} & \multicolumn{1}{c}{4C~08.23}& \multicolumn{1}{c}{4C~06.27}  \\
\hline
RA$_{\rm NVSS}$ (h:m:s)                                            &06:56:41.84                  &06:55:48.36                &06:59:17.97                  &06:58:46.96               &06:56:01.03                  &06:55:59.37   \\
RA$_{\rm FAST}$ (h:m:s)                                             &06:56:42.50                  &06:55:49.37                &06:59:18.47                 &06:58:46.86               &06:56:01.86                   &06:56:00.24  \\
DEC$_{\rm NVSS}$ ($\degr$:$\arcmin$:$\arcsec$)  &$+$11:55:18.1              &$+$10:42:58.8           &$+$08:13:31.8              &$+$08:20:20.9          &$+$08:34:07.1               &$+$06:39:00.8 \\
DEC$_{\rm FAST}$ ($\degr$:$\arcmin$:$\arcsec$)   &$+$11:55:13.4              &$+$10:42:53.2           &$+$08:13:32.6              &$+$08:20:28.9          &$+$08:34:13.9              &$+$06:39:18.8 \\
Offset ($\arcsec$)                                                     &10.8                               &15.9                            & 7.5                               & 8.1                             &14.1                               &22.2 \\
$S_{\rm nvss}$ (mJy)                                                    & 374$\pm$75               &910$\pm$180           &908$\pm$180              &374$\pm$75             &679$\pm$140             &455$\pm$91 \\
$S_{\rm FAST}$ (mJy)                                                   & 364$\pm$19               &877$\pm$45              &775$\pm$39               &381$\pm$20              &683$\pm$35               &414$\pm$21 \\
$\alpha_{\rm SPECFIND}$                                             &$-$1.03$\pm$0.05      &$-$0.78$\pm$0.05     &$+$0.03$\pm$0.04    &$-$1.02$\pm$0.06     &$-$0.94$\pm$0.04     &$-$0.73$\pm$0.04 \\
$\alpha_{\rm FAST}$                                                    &$-$1.13$\pm$0.19      &$-$0.97$\pm$0.19     &$-$0.16$\pm$0.19     &$-$1.16$\pm$0.19     &$-$1.17$\pm$0.19     &$-$0.98$\pm$0.19 \\
\hline
\end{tabular}
\end{table*}

\subsubsection{Brightness temperature}

Instead of the instrumental dependent antenna temperature $T_{A}$, the
main-beam brightness temperature $T_{B}$ is used when analyzing
astronomical objects. The relation between $T_{A}$ and $T_{B}$ can be
written as $T_{A} / T_{B} = \Omega_{mb} / \Omega_{A}$, where
$\Omega_{mb}$ and $\Omega_{A}$ denote the main-beam and beam solid
angle, respectively. $\Omega_{mb}$ can be estimated via $\Omega_{mb} =
$ 1.13 $\times~\theta^2$ for a Gaussian beam, where $\theta$ is the
HPBW in radian. $\Omega_{A}$ can be calculated as $\Omega_{A} =
\lambda^2 \times S_{\lambda} / ( 2k \times 10^{23} \times T_{A})$,
where $\lambda$ represents the observing wavelength in centimeter, k
is the Boltzmann constant 1.38 $\times 10^{-16} erg/K$, and
$S_{\lambda} / T_{A}$ is the reciprocal of the gain in Jy/K
T$_{A}$. By combining the above expressions, the relation between
$T_{A}$ and $T_{B}$ at a given observing wavelength $\lambda$ can be
simplified as:
\begin{equation}
  T_{A}/T_{B} = 2.647\ G \ \theta^{2}/\lambda^{2} \\
  \label{eq:LebsequeIII}
\end{equation}
where the units of the gain $G$, the HPBW $\theta$ and the observing
wavelength $\lambda$ are used in the units of $\rm K\ T_{A}/ Jy$,
arcmin, and centimeter, respectively. The ratio of $T_{A} / T_{B}$ is
also known as the main-beam efficiency. As for the FAST gain, we
present the FAST main-beam efficiency also in four groups (see the
bottom panel of Fig.~\ref{performance}). It shows that the FAST L-band
main-beam efficiency ranges from $\sim$0.65 to $\sim$0.85. The value
increases with the observing frequencies in general. This is unusual
but not unique. The same trend is seen in the 1.28-1.75~GHz survey
conducted by the 25.6-m telescope of the Dominion Radio Astrophysical
Observatory \citep{Wolleben21}. These factors, which consider the HPBW
and gain of different beams at different observing frequencies, are
used to convert the FAST image data into $T_{B}$ units.

\subsection{Data processing of FAST observations}

The FAST L-band receiver covers a frequency range of 1.0~GHz to
1.5~GHz, and the band is split into 65\,536 channels in the digital
backend. The first and the last $\sim$35~MHz of the entire bandwidth
were not used because of the low system gain. The central continuous
1.145-1.305~GHz band was heavily contaminated by radio
frequency interference (RFI) during the observations and the data in
this frequency range were discarded. The Galactic \ion{H}{I} emission,
centered at 1420.41~MHz covering about 70 channels ($\sim$0.5~MHz) and
a narrow band centered at 1.09~GHz ($\sim$3~MHz wide) affected by the
civil aviation were both masked. To eliminate the remaining accidental
RFI, two methods were tried. The asymmetrically re-weighted penalized
least squares smoothing \citep[ArPLS,][]{Baek15} has been demonstrated
to be very effective in flagging RFI \citep{Zeng21}.  The second is
simply the iteration of clipping signals above the $3\sigma$ r.m.s
level of the band value in the frequency domain. The ArPLS method
results in a relatively high percentage of RFI elimination, however,
consumes much more computing time. Therefore the second method was
adopted in this work, which achieved good results with much less
computing time. After RFI mitigation, about a half of the total
500~MHz bandwidth was left. For the in-band spectral-index
determination, we split the available bandwidth into five
sub-bands. They are centered at 1.065~GHz, 1.118~GHz, 1.334~GHz,
1.391~GHz, and 1.443~GHz with a bandwidth of approximately 50~MHz
each. For such a small frequency span, the data within each sub-band
were combined neglecting the small difference in angular resolution.

\citet{Sun21} made the FAST radio continuum polarization calibration
based on the polarization characteristics of the injected reference
signal. The differential gain between the Stokes $I$ and $Q$, and the
differential phase between the Stokes $U$ and $V$ were corrected. The
method was demonstrated to work well with the FAST radio continuum
data. The same scheme is applied in this work. The detailed procedure
can be referred to \citet{Sun21} and will not be repeated here.

\begin{figure*}[t]
  \centering
  %\hspace{-2.5cm}
  \includegraphics[width = 0.97 \textwidth]{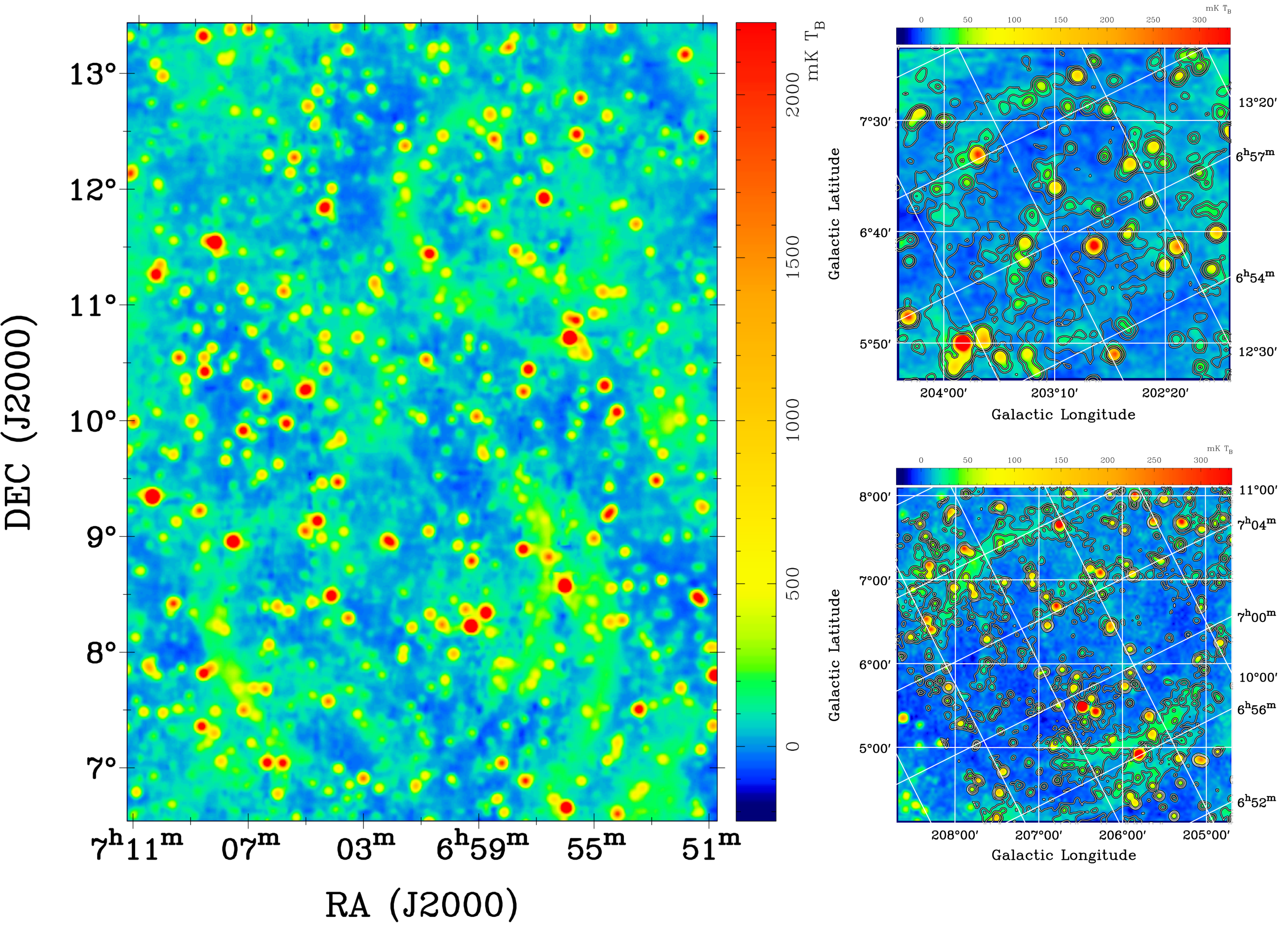}
  \caption{{\it Left panel:} FAST L-band total-intensity image for
    G203.1+6.6 (centered at $\rm RA = 06^h57^m08.9^s, DEC =
    11\degr42\arcmin27.9\arcsec$) and G206.7+5.9 (centered at $\rm RA
    = 07^{h}01^{m}18.8^{s}$, $\rm DEC = 08\degr13\arcmin11.6\arcsec$)
    with an angular resolution of 4$\arcmin$. {\it Right panel:}
    Effelsberg $\lambda$11 \ cm total-intensity images of G203.1+6.6
    (top) and G206.7+5.9 (bottom). FAST L-band total-intensity
    contours at levels of 2$^{n - 1}$ $\times$ 99~mK\ T$_{B}$ (n = 1,
    2, 3) are overlaid onto the Effelsberg data for comparison.}
\label{Fig_I}
\end{figure*}

\subsection{Verification of FAST data processing}

Recently, two bright SNRs G74.0$-$8.5 \citep[Cygnus Loop,][]{Sun21,
  Sun22} and G166.0+4.3 \citep[VRO 42.05.01,][]{Xiao22} were studied
based on the FAST L-band radio continuum observations. The
data-processing pipelines were separately developed. The Cygnus Loop
and VRO 42.05.01 are both well-known SNRs, whose flux density and
spectrum can be readily compared with the previously published
values. However, the two objects G203.1+6.6 and G206.7+5.9 presented
in this paper are investigated for the first time. Therefore, the
data-processing procedure and the correctness of the derived in-band
spectral index should be verified prior to the study. Strong
point-like sources from the NVSS \citep{Condon98} were taken for this
purpose. The image containing G203.1+6.6 and G206.7+5.9 was retrieved
from the NVSS database\footnote{https://www.cv.nrao.edu/nvss/} and
compared with the FAST image. The sources detected by FAST and cannot
be further resolved into multiple components by NVSS were required.
Finally, six strong sources, 4C~11.24 and 4C~10.21 within the field of
G203.1+6.6, and ICRF J065917.9+081330, 4C~08.24, 4C~08.23, and
4C~06.27 within G206.7+5.9 were chosen. The positional accuracy of
FAST was verified by the comparison between the coordinates obtained
from the 2D Gaussian fits to these selected sources in the FAST image
and those extracted from the NVSS source catalog. The differences are
7.5$\arcsec$ to 22.2$\arcsec$ (see Table.~\ref{tab:tab1}), which is
acceptable for the FAST HPBW of 3$\arcmin$. The NVSS 1.4-GHz (mean of
1364.9~MHz \& 1435.1~MHz) flux densities of the sources are listed
against the average values measured by FAST at 1.391~GHz and
1.443~GHz, which shows good consistency between the two sets of data.
The FAST in-band spectral indices were derived from the flux densities
determined at the five sub-bands. The uncertainty of the total flux
density is about 5\%, which consists of the instability of the
injected reference signals, the effect from the pointing accuracy
\citep{Jiang20}, the uncertainties of the main-beam efficiency, and
the Gaussian fits. For comparison, we fit the flux-density data
extracted from the third version of SPECFIND \citep{Stein21}. The two
sets of spectral indices are listed in the last two rows of
Table~\ref{tab:tab1}. The spectral indices based on the SPECFIND data
are derived over a much wider frequency range, while the largest
frequency span of FAST is only $\sim$380~MHz. The FAST in-band spectra
are slightly steeper, but still agree with the SPECFIND results within
uncertainties. We noticed that a better consistency between the two
sets of results is achieved when the data in SPECFIND below 100~MHz or
200~MHz are omitted, especially for the sources 4C~11.24, 4C~10.21,
and 4C~08.23. This may result from the absorption that flattens the
spectra at the low-frequency end.

Based on the results shown above, the FAST data processing procedures
are well proved and can be used for further radio continuum study.

\begin{figure*}
\centering
\resizebox{0.24\textwidth}{!}{\includegraphics[angle = -90]{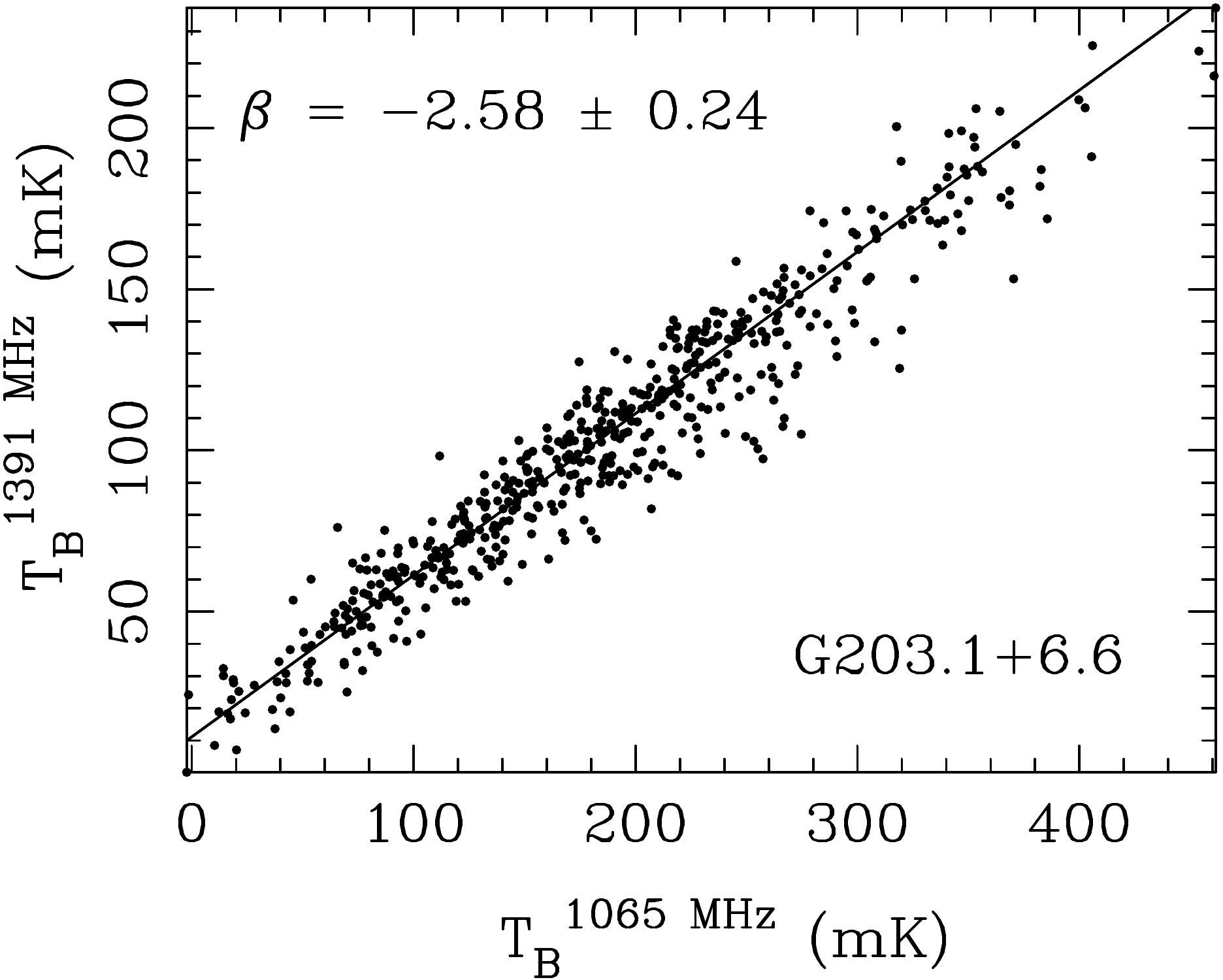}}
\resizebox{0.24\textwidth}{!}{\includegraphics[angle = -90]{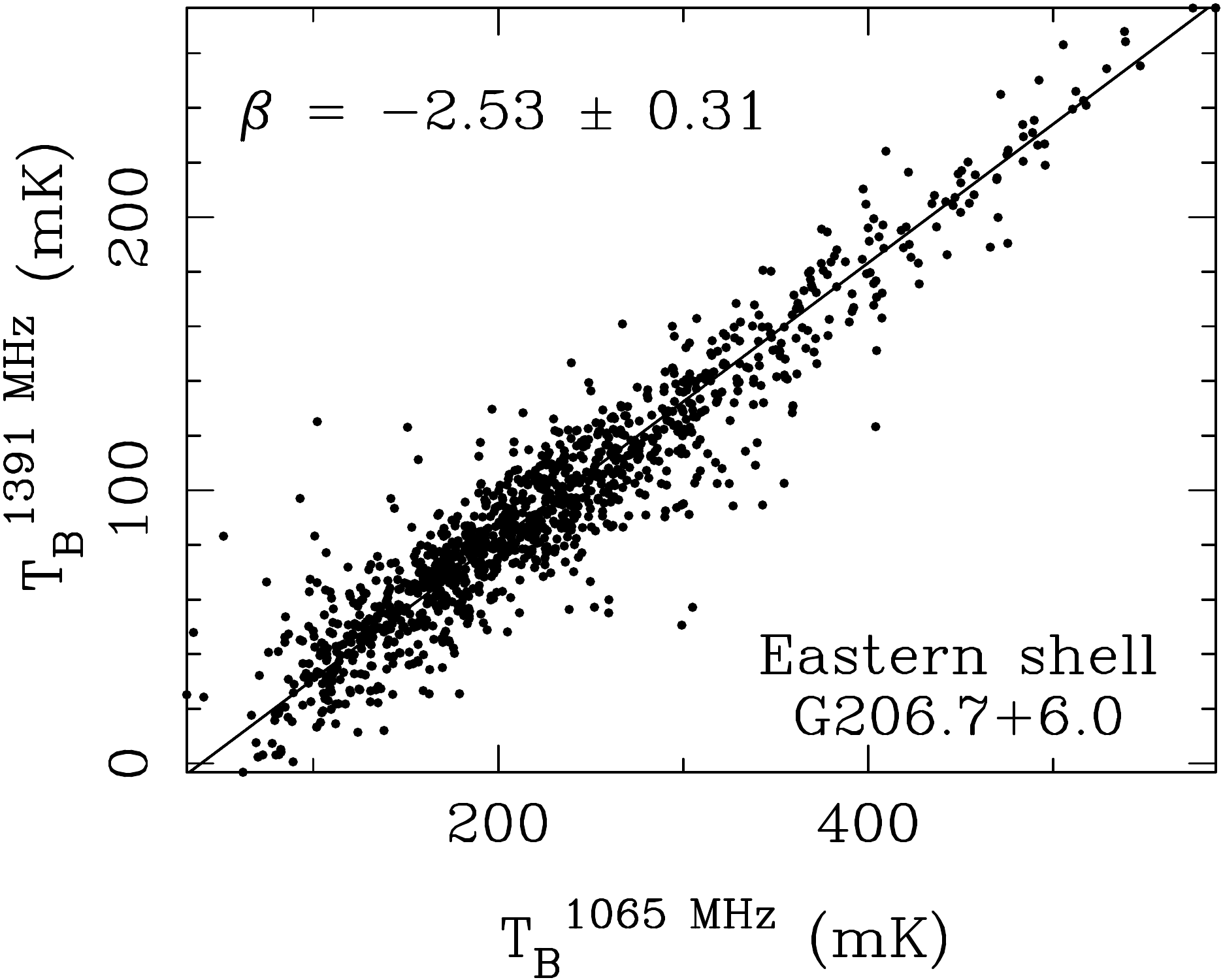}}
\resizebox{0.24\textwidth}{!}{\includegraphics[angle = -90]{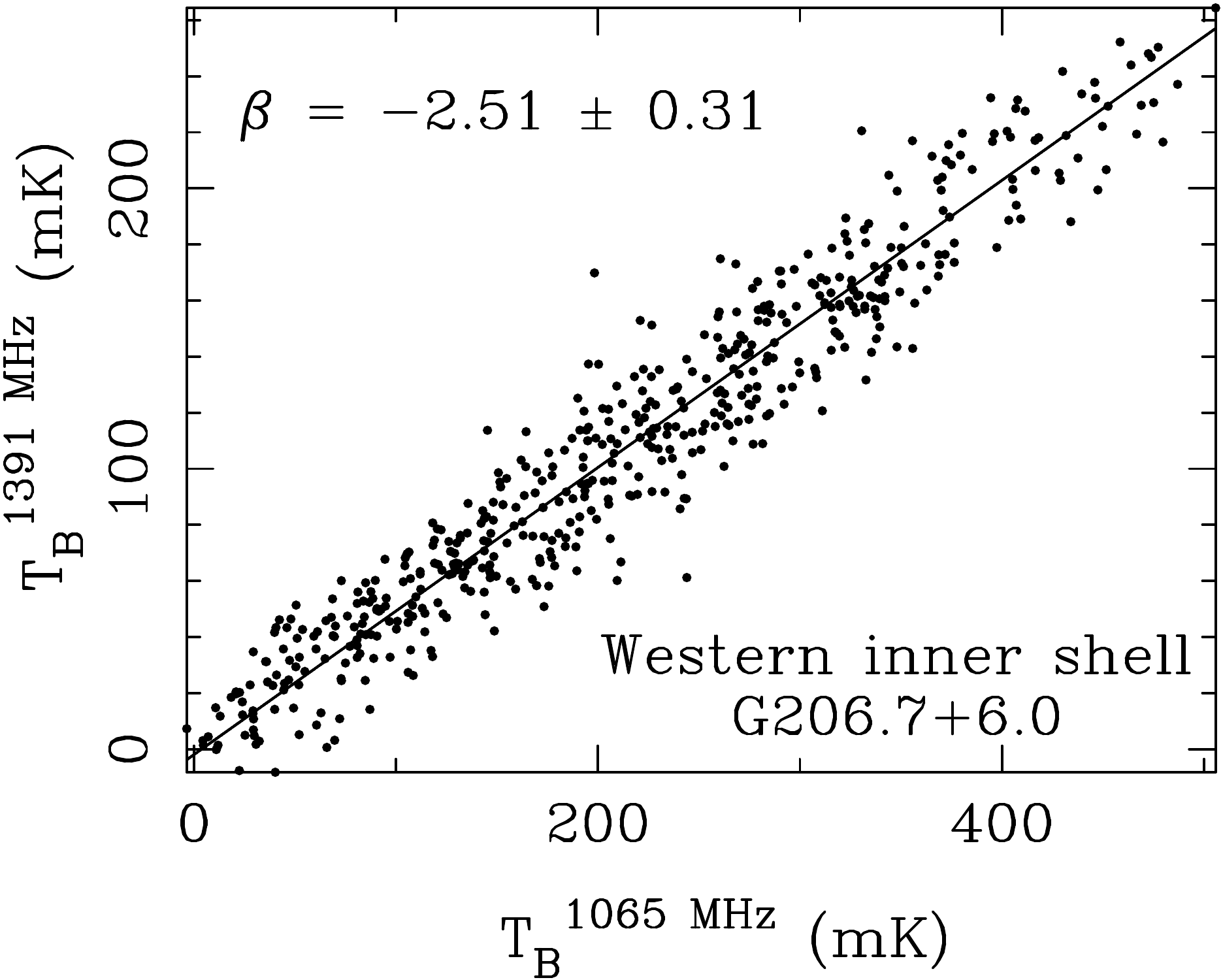}}
\resizebox{0.24\textwidth}{!}{\includegraphics[angle = -90]{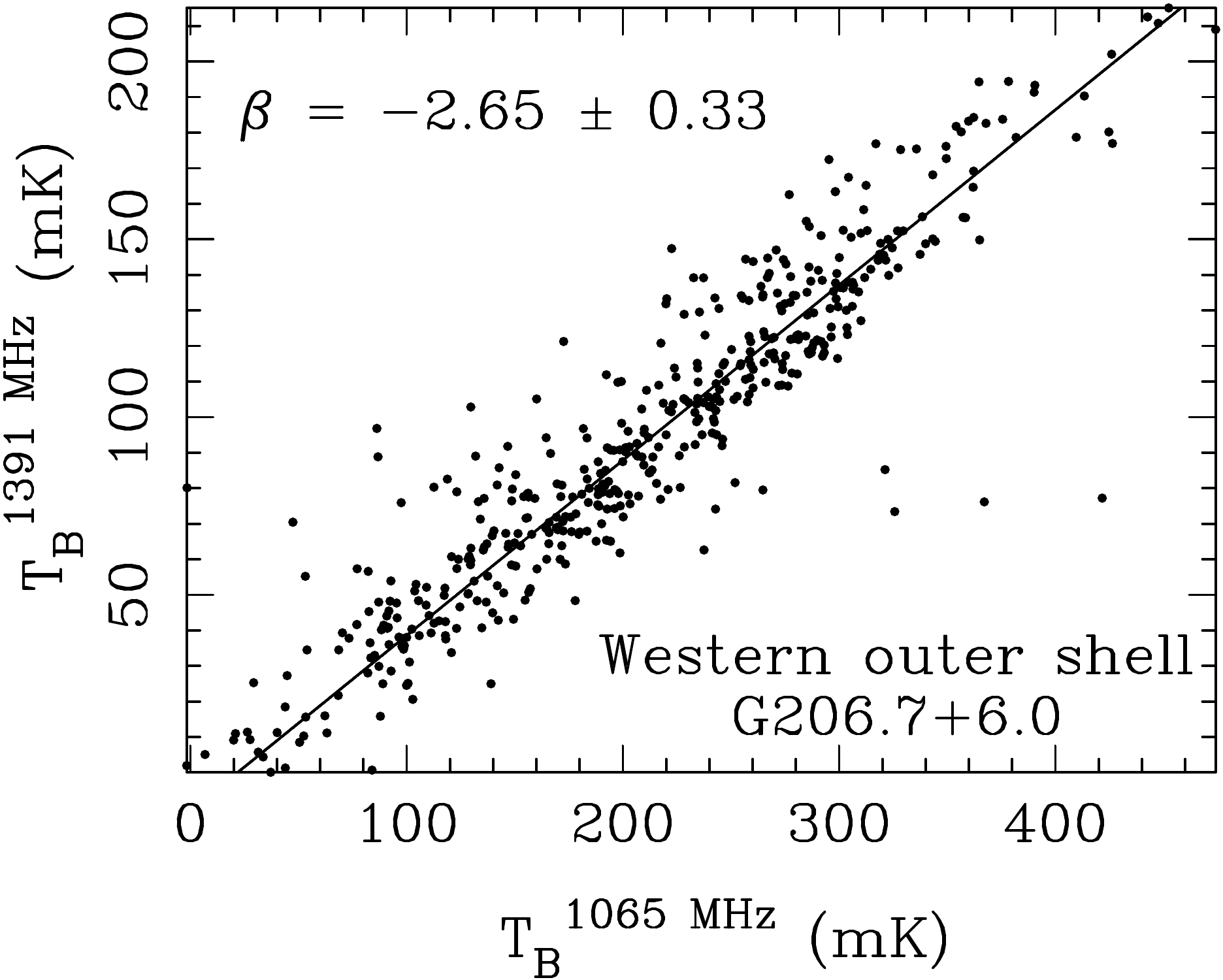}} \\
\resizebox{0.24\textwidth}{!}{\includegraphics[angle = -90]{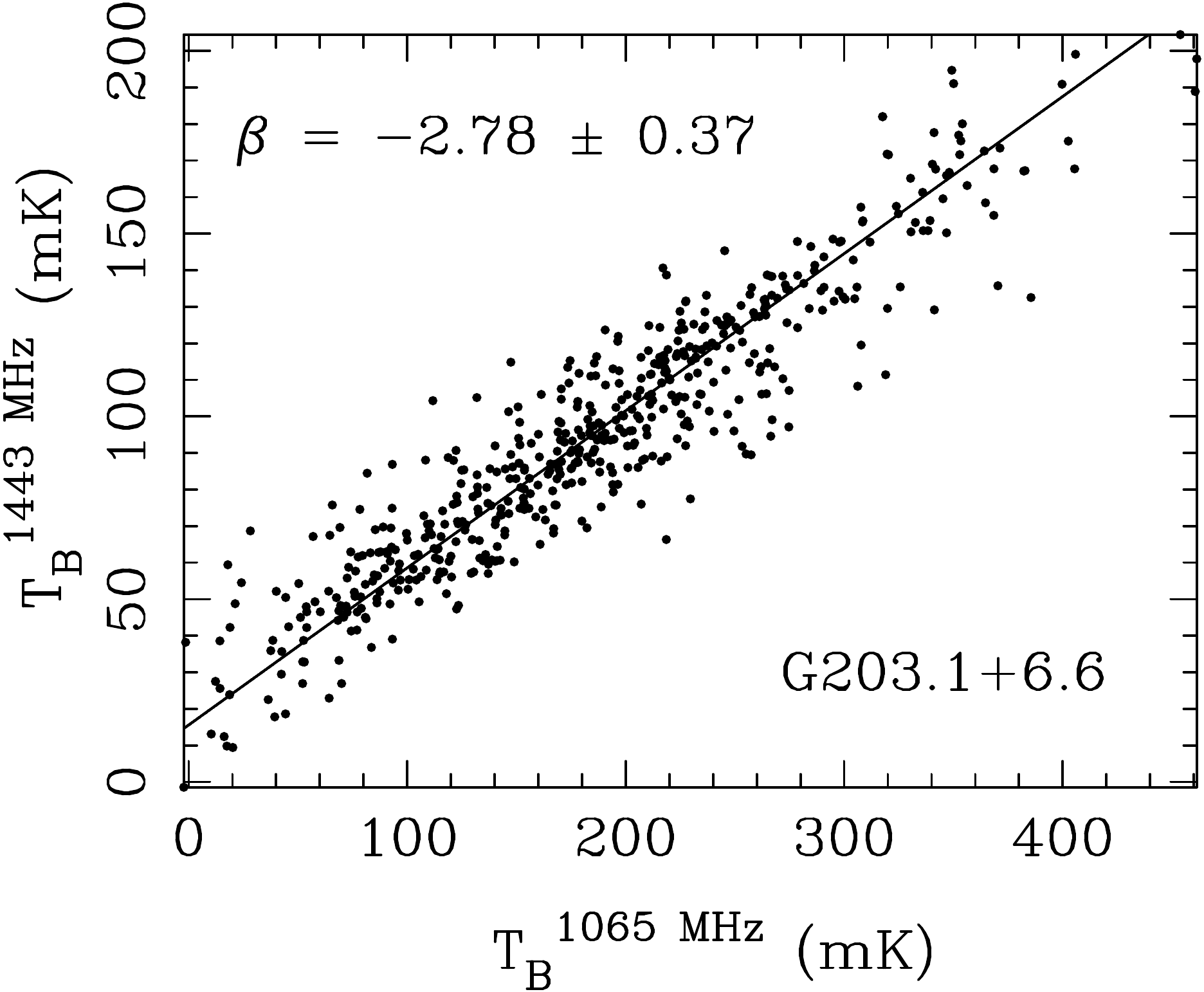}}
\resizebox{0.24\textwidth}{!}{\includegraphics[angle = -90]{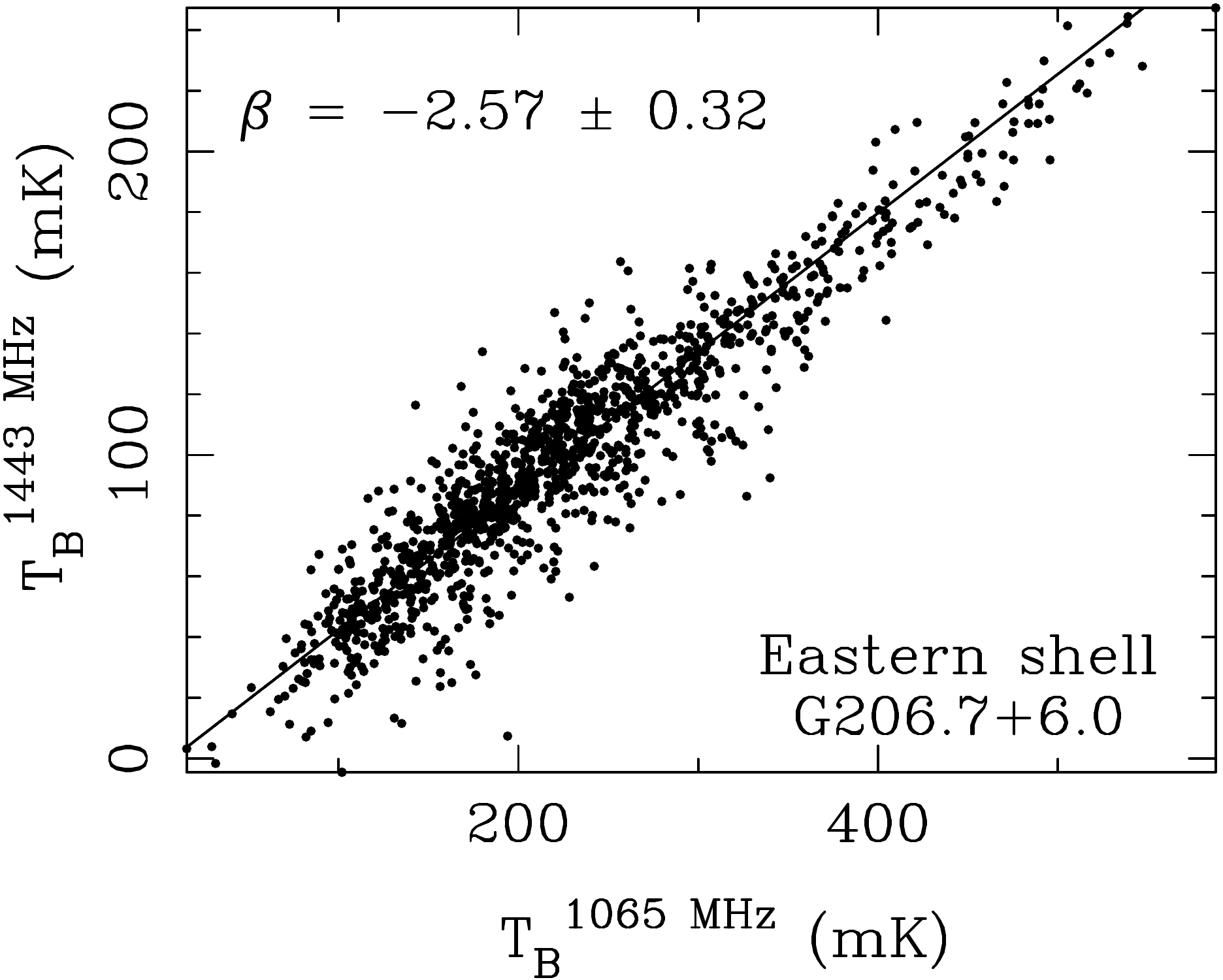}}
\resizebox{0.24\textwidth}{!}{\includegraphics[angle = -90]{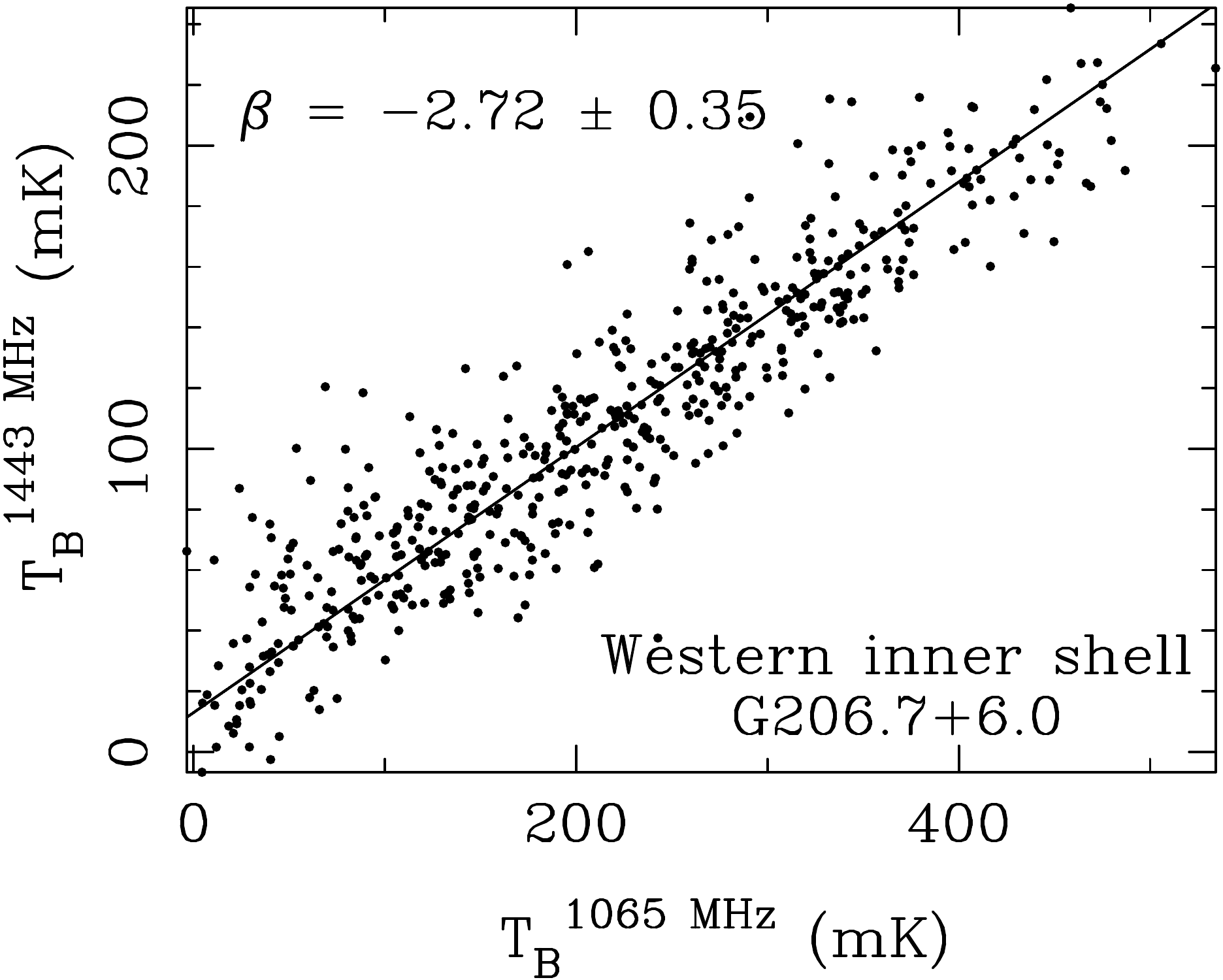}}
\resizebox{0.24\textwidth}{!}{\includegraphics[angle = -90]{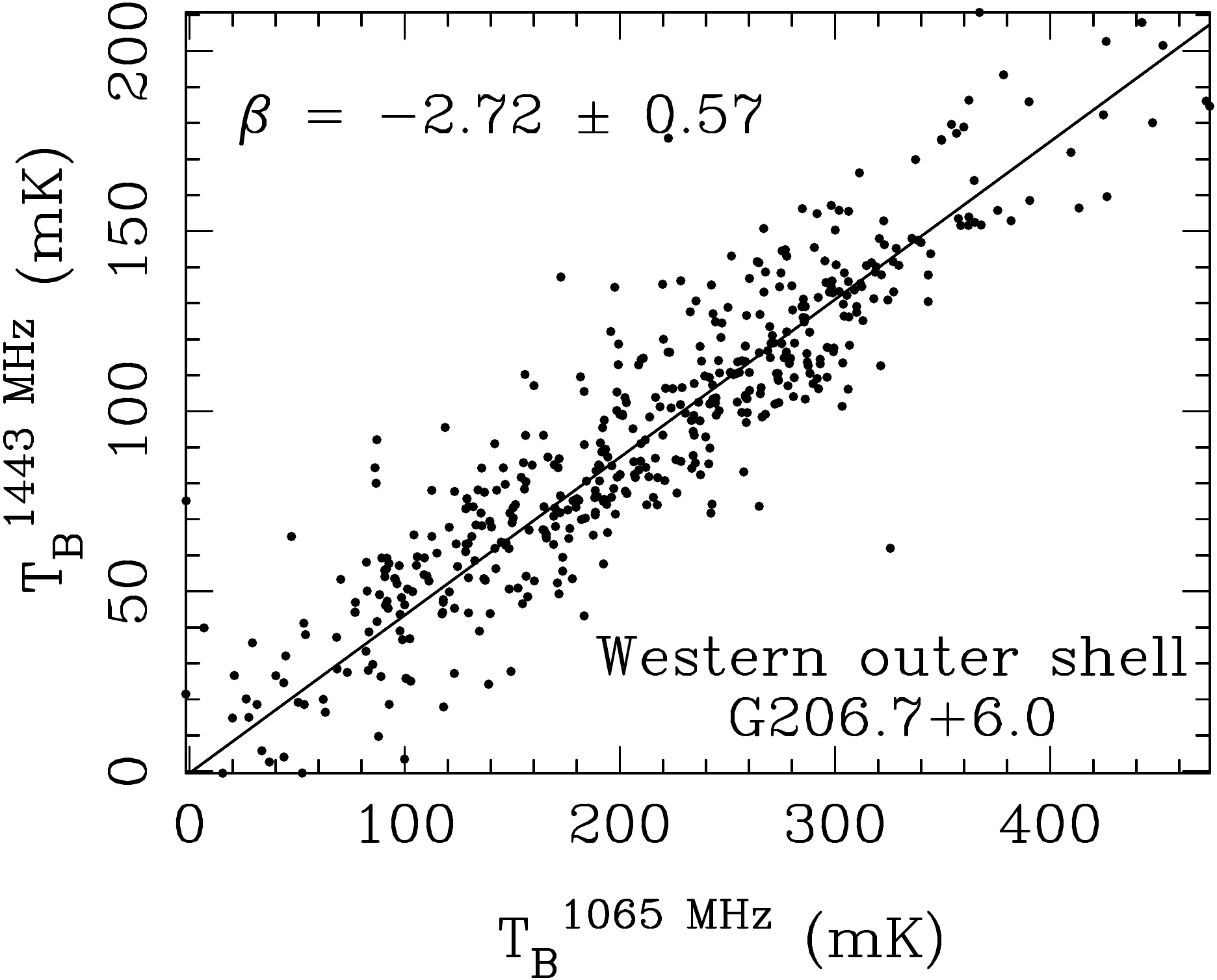}} \\
\caption{Temperature-versus-Temperature plots for deriving the
  brightness-temperature spectral index for G203.1+6.6, and for the
  three shells of G206.7+5.9.}
\label{TT-plot}
\end{figure*}

\subsection{Effelsberg $\lambda$11\ cm observations}

We complemented the FAST L-band observations with the archival
$\lambda$11\ cm observations by the Effelsberg 100-m telescope in
2006. \citet{Foster13} presented SNR observations conducted at the
same period and gave a detailed description of the Effelsberg
$\lambda$11\ cm receiver, the backend, and also the observing method
and reduction procedures. We therefore give a brief summary of the key
parameters. The central observing frequency was 2639~MHz and the
bandwidth was 80~MHz. The HPBW of the 100-m telescope at this
frequency band is about 4$\farcm$4, similar to that of FAST at
L-band. The IF-polarimeter has eight channels with 10-MHz bandwidth
each and a broadband 80-MHz channel. All channels have their Stokes
$I$, $U$, and $Q$. The calibration was based on 3C~286 assuming a flux
density of 10.4~Jy, with 9.9\% of linear polarization and a
polarization position angle of 33$\degr$. G203.1+6.6 and G206.7+5.9
were observed separately and ten maps were obtained for each. The data
of the 80-MHz channel were used when RFI was low, otherwise the average
data from the seven narrow-band channels were used after the lowest
frequency channel (2599~MHz to 2609~MHz) was omitted because of strong
RFI.

\section{Results}
\label{sect:results}

We identify G203.1+6.6 and G206.7+5.9 as SNRs according to their
non-thermal nature by showing steep radio continuum spectra and
polarized emission. We also try to estimate the distances of these two
objects based on the morphological correlation between radio continuum
and \ion{H}{I} emission.

\subsection{Total-intensity map and spectrum determination}

We present the total-intensity $I$ map of G203.1+6.6 and G206.7+5.9
observed by FAST at L band in Fig.~\ref{Fig_I}. The image is a
combination of the five sub-band maps after they were convolved to a
common angular resolution of 4$\arcmin$. The RA and Dec directional
scans were weaved together by using the ``basket-weaving'' technique
\citep{Emerson88}. The rms noise of the image is about 14~mK
$T_{B}$. G203.1+6.6 shows a large arc which spans about 2$\fdg$5 and
opens to the west. At the center, a smaller and dim ring-like
structure of about 1$\degr$ can also be identified. However, the weak
and fuzzy emission from some parts of the ring is difficult to
discriminate because of the presence of numerous unresolved
extra-galactic sources. G206.7+5.9 is a large object of about 3$\fdg$5
in diameter with a classical bilateral structure. It has one shell in
the east and two shells in the west (in the Equatorial
coordinates). Such an appearance could be a result of SNR shock front
expanding perpendicular to a well-ordered large-scale magnetic field
\citep[e.g.][]{VanderLaan62, Gaensler98}. Therefore the shells may
indicate the orientations of the local magnetic field.

The Effelsberg $\lambda$11\ cm total-intensity images of G203.1+6.6
and G206.7+5.9 are shown in the right panels of Fig.~\ref{Fig_I}. The
angular resolution is 4$\farcm$4. Unlike the FAST observations made in
the Equatorial coordinates, Effelsberg observations were conducted in
the Galactic frame. We therefore overlaid the RA-DEC coordinates and
FAST total-intensity contours onto the Effelsberg images for
reference. As expected, the total-intensity emission structures
revealed by FAST and Effelsberg resemble each other under similar
angular resolution.

\begin{figure*}[!t]
  \centering
    \includegraphics[height=0.44\textwidth,angle=-90]{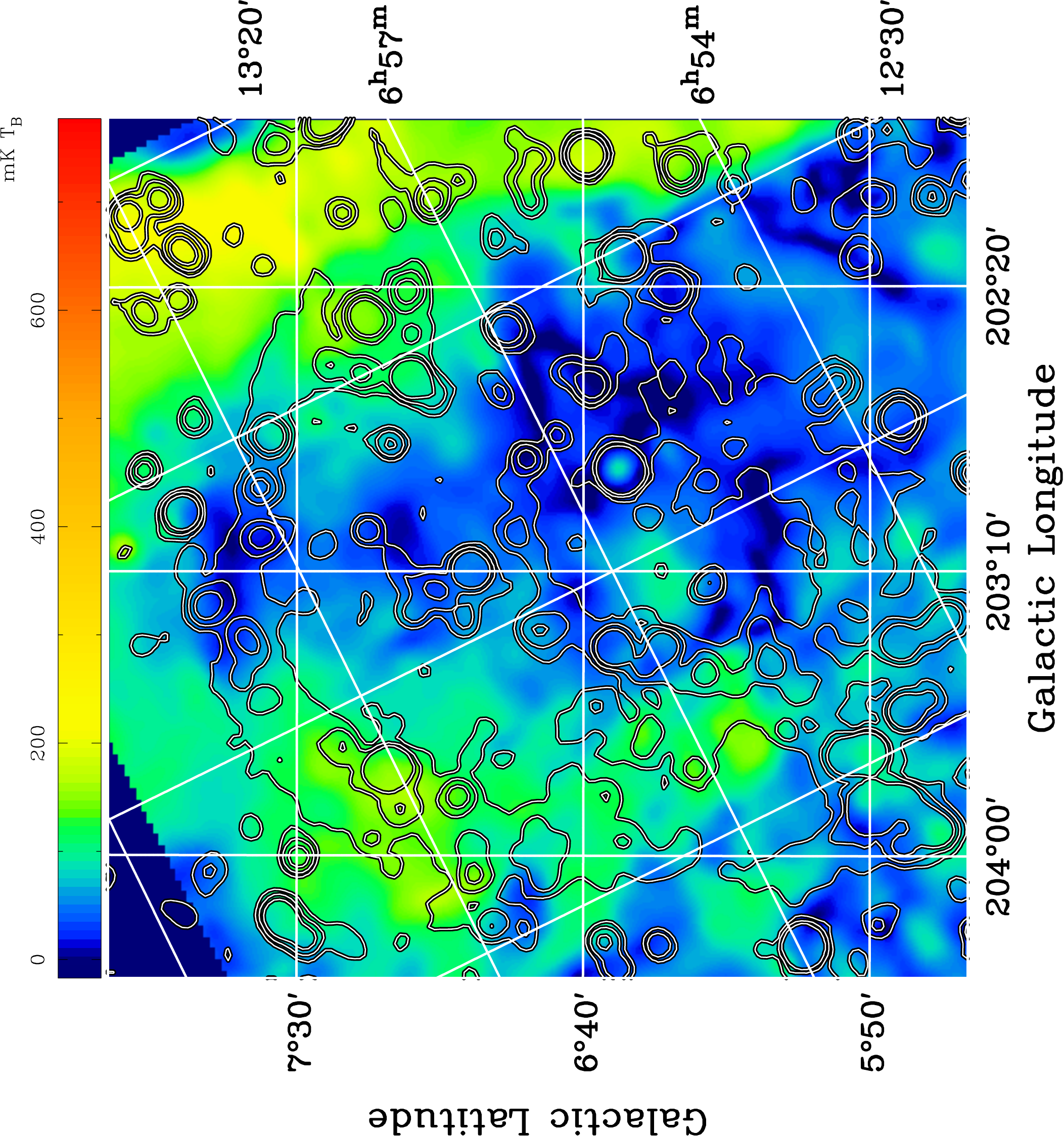}
    \includegraphics[height=0.44\textwidth,angle=-90]{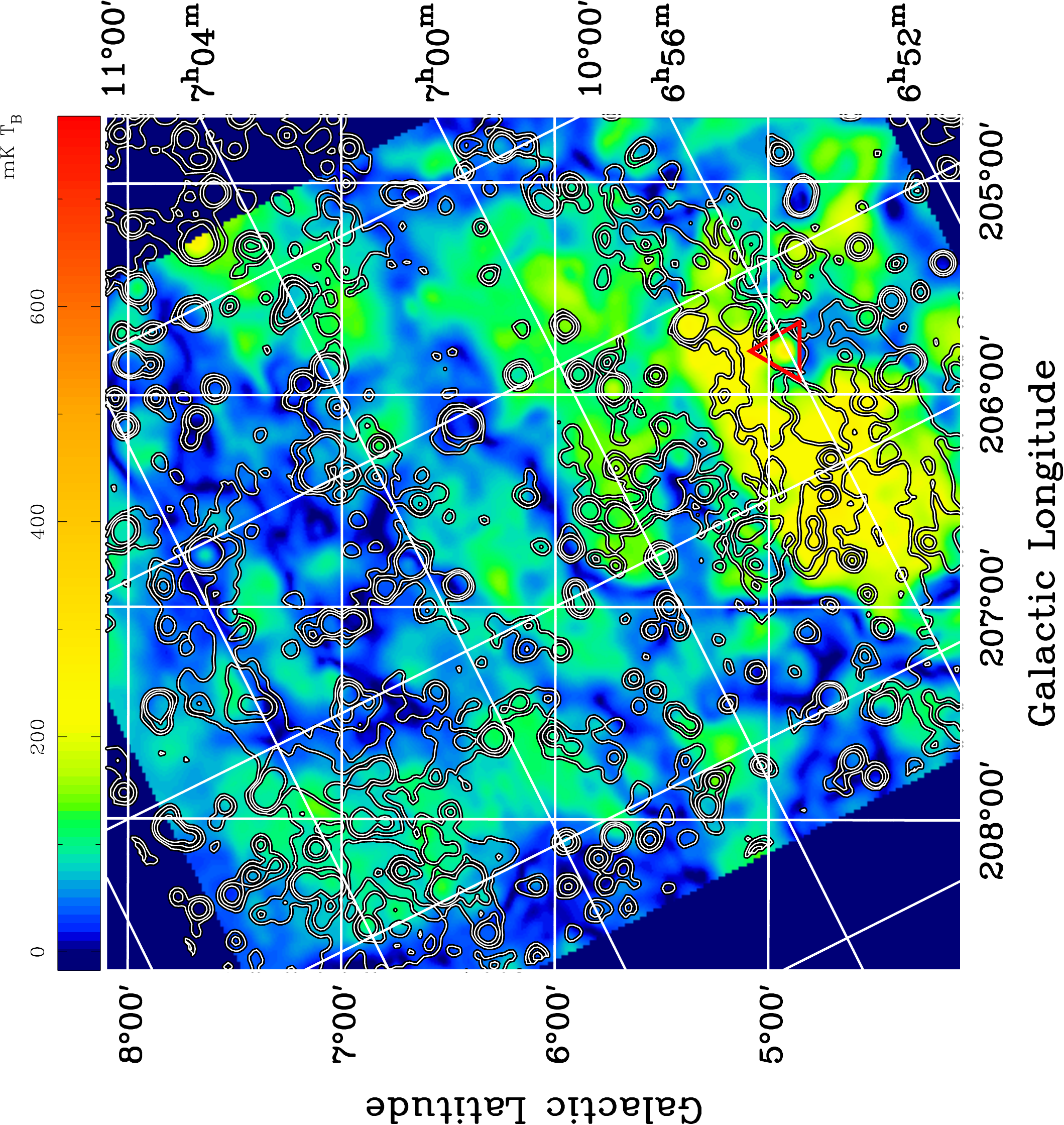}\\
    \includegraphics[height=0.44\textwidth,angle=-90]{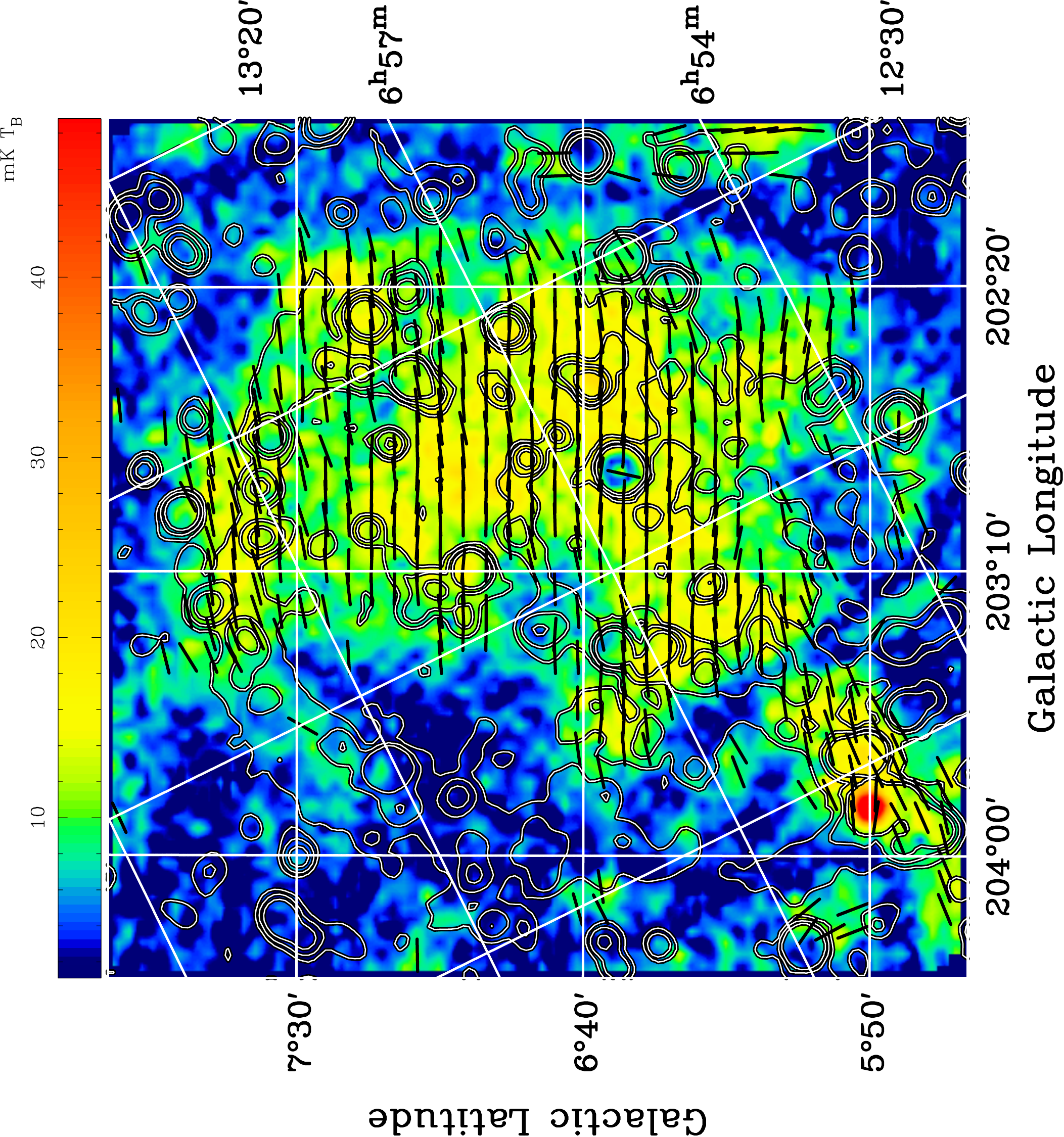}
    \includegraphics[height=0.44\textwidth,angle=-90]{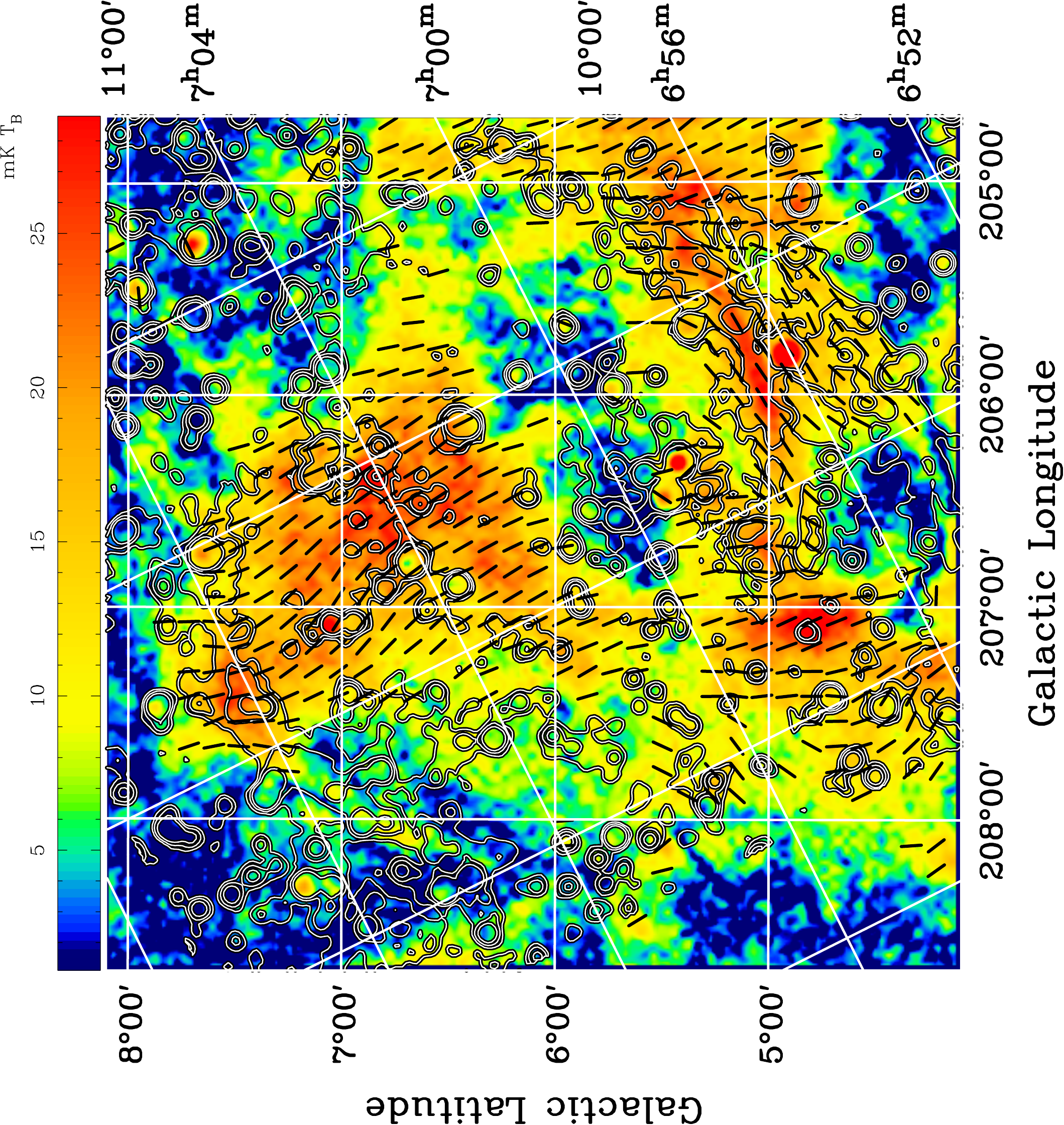} \\
  \caption{{\it Upper panels:} FAST 1.443-GHz polarization-intensity
    ($PI$) images of G203.1+6.6 (left) and G206.7+5.9 (right) with an
    angular resolution of 5$\arcmin$. The red triangle marked in
    G206.7+5.9 indicates the source 4C~08.23 with a measured $RM$ of
    26$\pm$8~rad\ m$^{-2}$. The contours of total intensities measured
    by FAST (the same as in Fig.~\ref{Fig_I}) illustrate the shell
    structures of the two SNRs. {\it Lower panels:} the same as in the
    {\it upper panels}, but for the Effelsberg $\lambda$11\ cm
    polarization intensity ($PI$) at their original angular resolution
    of 4$\farcm$4. Overlaid bars indicate the magnetic field
    orientations in case Faraday rotation can be neglected.}
\label{Fig_11PI}
\end{figure*}

Spectral index is a key parameter to distinguish between
optically-thin \ion{H}{II} regions ($\alpha \sim -0.1$, $S_{\nu} \sim
\nu^\alpha$) and non-thermal synchrotron emitting shell-type SNRs,
most of which have a spectral index around $\alpha \sim -0.5$
\citep[e.g][]{Dubner15} in their adiabatic evolution phase.  The
Temperature-Temperature plots, namely TT-plots \citep{Turtle62},
provide a reliable tool to assess the spectral index. They are less
affected by base-level differences between the maps and therefore have
often been used when analyzing radio continuum maps. The brightness
temperatures of corresponding pixels observed at two different
frequencies are compared. A good linear relation implies the existence
of a power-law spectrum of the extended emission, while the slope
provides the brightness-temperature spectral index $\beta$, which
relates to the flux-density spectral index $\alpha$ as $\beta = -2 +
\alpha$. We performed the TT-plot analysis to the arc of G203.1+6.6
and the three shells of G206.7+5.9. The FAST radio maps at five
sub-bands were analyzed. First, we masked the un-related point-like
sources overlapping the shell structures based on the NVSS source
catalog \citep{Condon98}. To alleviate the uncertainties introduced by
using a narrow frequency separation, the data in the first sub-band
(1065~MHz) are plotted against those of the 4th (1391~MHz) and 5th
(1443~MHz) sub-band. The results are shown in Fig.~\ref{TT-plot}. For
G203.1+6.6, the spectral index is found to be $\beta \sim -2.6\pm0.2$
between 1065 and 1391~MHz, and $\beta \sim -2.8\pm0.4$ between
1065 and 1443~MHz. Both values agree with each other within
errors. Analyzing in the same way, we found a spectral index of $\beta
\sim -2.6$ for the eastern shell of G206.7+5.9, while the spectral
indices of the inner and outer shells in the west are $\beta \sim
-2.6$ and $\beta \sim -2.7$, respectively. These results provide
strong evidence that G203.1+6.6 and G206.7+5.9 are extended,
shell-type, synchrotron-emitting SNRs. We also tried to estimate the
brightness-temperature spectral index between the FAST L-band and the
Effelsberg $\lambda$11\ cm data. We got $\beta = -2.8\pm0.5$ for the
arc of G203.1+6.6, and $\beta \sim -2.7$ for all the three shells of
G206.7+5.9 with a large uncertainty. No reliable TT-plot results
were obtained for the central ring structure inside G203.1+6.6.

\subsection{Polarization}
\label{sect:pol}

Synchrotron radiation is intrinsically polarized. The absence of
polarization detection is often caused by depolarization, e.g. beam
depolarization and/or depth depolarization \citep{Burn66,
  Sokoloff98}. Beam depolarization occurs when varying polarized
emission regions smaller than the beam size are averaged within the
beam, while depth depolarization is caused by the mixture of
synchrotron emitting medium and Faraday rotating medium. The polarized
emission originated in different depths along the same line of sight
undergoes different Faraday rotations. The polarized emission cancels
each others when added together. Because Faraday rotation is larger at
lower observing frequencies, such as in L band, the FAST Stokes $U$
and $Q$ data of each sub-band cannot be simply combined as we did for
the total intensity. The sub-band data at the highest observing
frequency of 1.443~GHz were selected and the images were convolved to
an angular resolution of 5$\arcmin$ to increase the signal to noise
ratio. The polarization intensity $PI$ is calculated as $PI =
\sqrt{U^2 + Q^2 - \sigma^2_{U,Q}}$ \citep{Wardle74}, where
$\sigma_{U,Q}$ is the rms noise of about 8~mK $T_{B}$ measured in the
$U$ and $Q$ maps. We show the FAST 1.443-GHz polarization-intensity
($PI$) images of G203.1+6.6 and G206.7+5.9 in
Fig.~\ref{Fig_11PI}. Some weak polarized emission is seen within the
arc of G203.1+6.6 and in the shell regions of G206.7+5.9. However, it
is difficult to verify the true association. In case of beam
depolarization which is additionally introduced as we smoothed the
polarization data, the $U$, $Q$, and the resulting $PI$ images were
examined at their original resolution. The above result remains that
no un-ambiguous polarized emission that is associated to the SNRs can
be confirmed. The most pronounced feature shown in the FAST 1.4-GHz
polarization image is an extended polarized ``blob'' centered at ($l,
b$) $\sim$ (206$\fdg$4, 4$\fdg$6) in G206.7+5.9. It can be identified
in all of the five sub-bands. The ``blob'' overlaps both of the double
shells in the southwest of G206.7+5.9 and the gap in between (Galactic
coordinates, see in Fig.~\ref{Fig_11PI}).

Rotation measure ($RM$) is the integral of magnetic field along the
line of sight, weighted by electron density. By fitting the
polarization angles of the entire ``blob'' against the wavelength
square in the five sub-bands, a good linear relation was obtained and
the $RM$ can be estimated ($RM = \Delta PA / \lambda^2$) to be
29$\pm$4~rad\ m$^{-2}$. We estimated the $RM$ values of the shell
region and the gap region separately. No significant difference was
found, suggesting that the ``blob'' itself is likely a discrete
structure unrelated to G206.7+5.9.
  
To search for the imprints of magnetic fields of G203.1+6.6 and
G206.7+5.9, we compared the $RM$ values of the sources whose line of
sight pass and do not pass the two large structures. We retrieved $RM$
values of extra-galactic sources from the catalog of \citet{Xu14}
within a radius of 3$\degr$ to the center of G203.1+6.6 and 4$\degr$
for G206.7+5.9, which are large enough to cover the two objects with a
radius of about 75$\arcmin$ and 105$\arcmin$, respectively. Seventeen
sources with measured $RM$ values were found outside G203.1+6.6. The
average $RM$ weighted by 1/$\sigma_{RM}$ was found to be
39$\pm$25~rad\ m$^{-2}$. In the same way, a similar $RM$ value of
40$\pm$21~rad\ m$^{-2}$ was found for G206.7+5.9 by averaging 27
sources in its vicinity. A possible double source which has two
components located at ($l, b$) = ($203\fdg75, 7\fdg25$) and ($l, b$) =
(203$\fdg$74, 7$\fdg$23) coincides in the arc region of
G203.1+6.6. The measured $RM$ values for the double source are
57.8$\pm$12.7~rad\ m$^{-2}$ and 67.5$\pm$14.2~rad\ m$^{-2}$,
respectively. However, with large uncertainties of the $RM$ values
both inside and outside the arc of G203.1+6.6, it is difficult to
claim a clear $RM$ excess caused by the SNR. The case is the same for
G206.7+5.9. The $RM$ values overlapping the north-eastern shell and in
the central region of G206.7+5.9 were found to be
51$\pm$10~rad\ m$^{-2}$ and 66$\pm$29~rad\ m$^{-2}$. They seem to be
larger but are still consistent with the reference value of
40$\pm$21~rad\ m$^{-2}$ within the errors. $RM$ changes introduced by
the SNRs depend on the view angles with respect to the magnetic
fields. The change could be significant through a compression to the
magnetic fields running parallel to the line of sight, because the
front and the rear parts of the SNR contribute the $RM$ with the same
sign. The change, however, may be insignificant when the line of sight
is perpendicular to the magnetic fields, since the signs of the $RM$s
become reverse in this case. Therefore, the $RM$ comparison shown
above implies that the magnetic field compressed by G203.1+6.6 and
G206.7+5.9 might predominantly run perpendicular to the line of sight
and contributes no significant $RM$. The source 4C~08.23 (indicated by
the red triangle in Fig.~\ref{Fig_11PI}) is close to the ``blob'' and
has a $RM$ value of 26$\pm$8~rad\ m$^{-2}$ \citep{Van11}. It is the
only $RM$ source found in the southwest of G206.7+5.9. The $RM$ value
of 4C~08.23 is smaller than those for the other sources
($\sim$55~rad\ m$^{-2}$) seen through G206.7+5.9, but agrees well with
$RM$ = 29 rad\ m$^{-2}$ derived for the ``blob''. Another $RM$ source
located at ($l, b$) = (205$\fdg$39, 4$\fdg$18) outside the
south-western double shell of G206.7+5.9 has a measured $RM$ of
17.9$\pm$10.6~rad\ m$^{-2}$. These three similar and smaller $RM$
values of the blob and through the Galaxy indicate that the $RM$
contribution from the background may be very small in this area.

\begin{figure*}[t]
\centering
\includegraphics[width=0.32\textwidth]{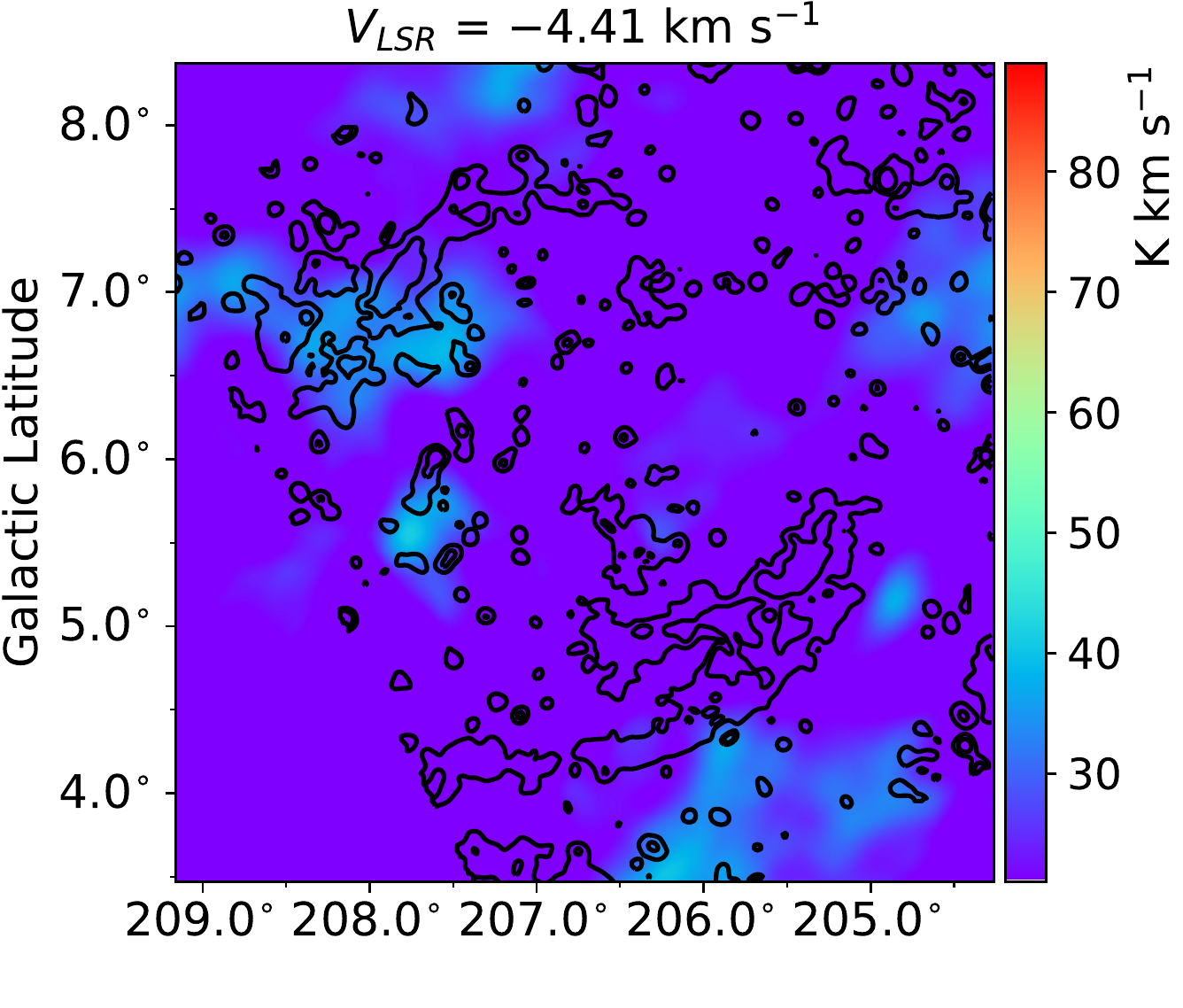}
\includegraphics[width=0.32\textwidth]{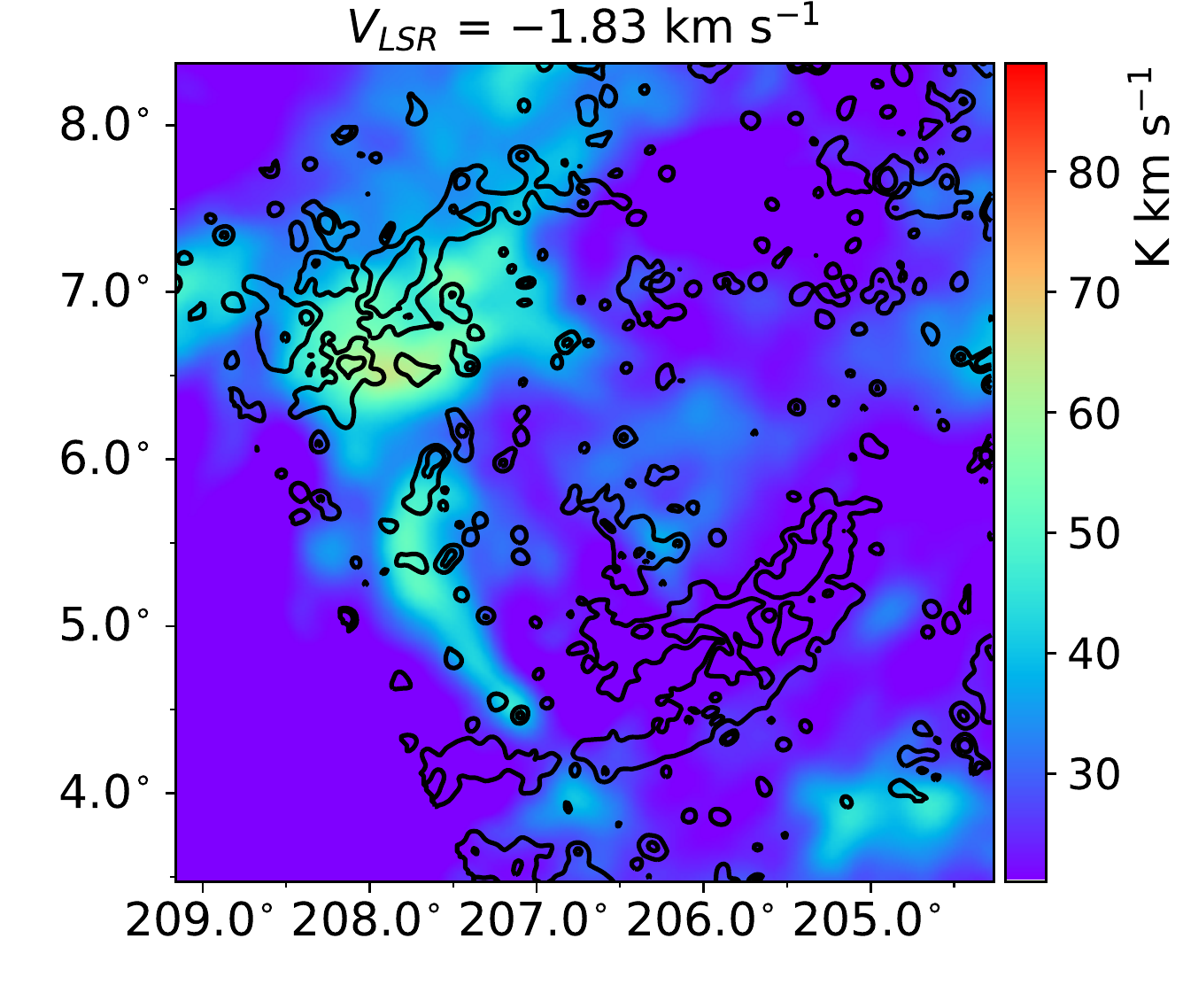}
\includegraphics[width=0.32\textwidth]{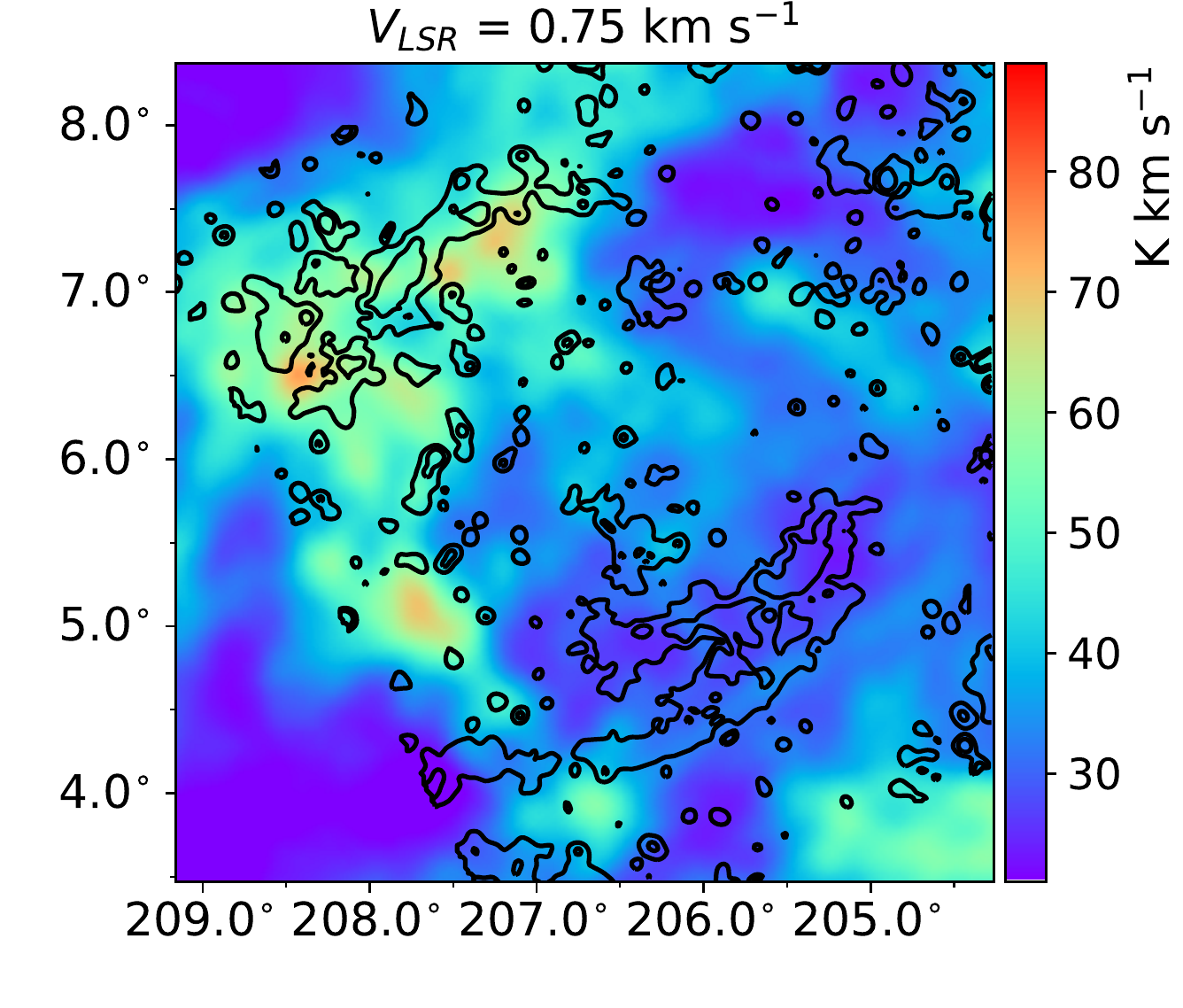}\\
\includegraphics[width=0.32\textwidth]{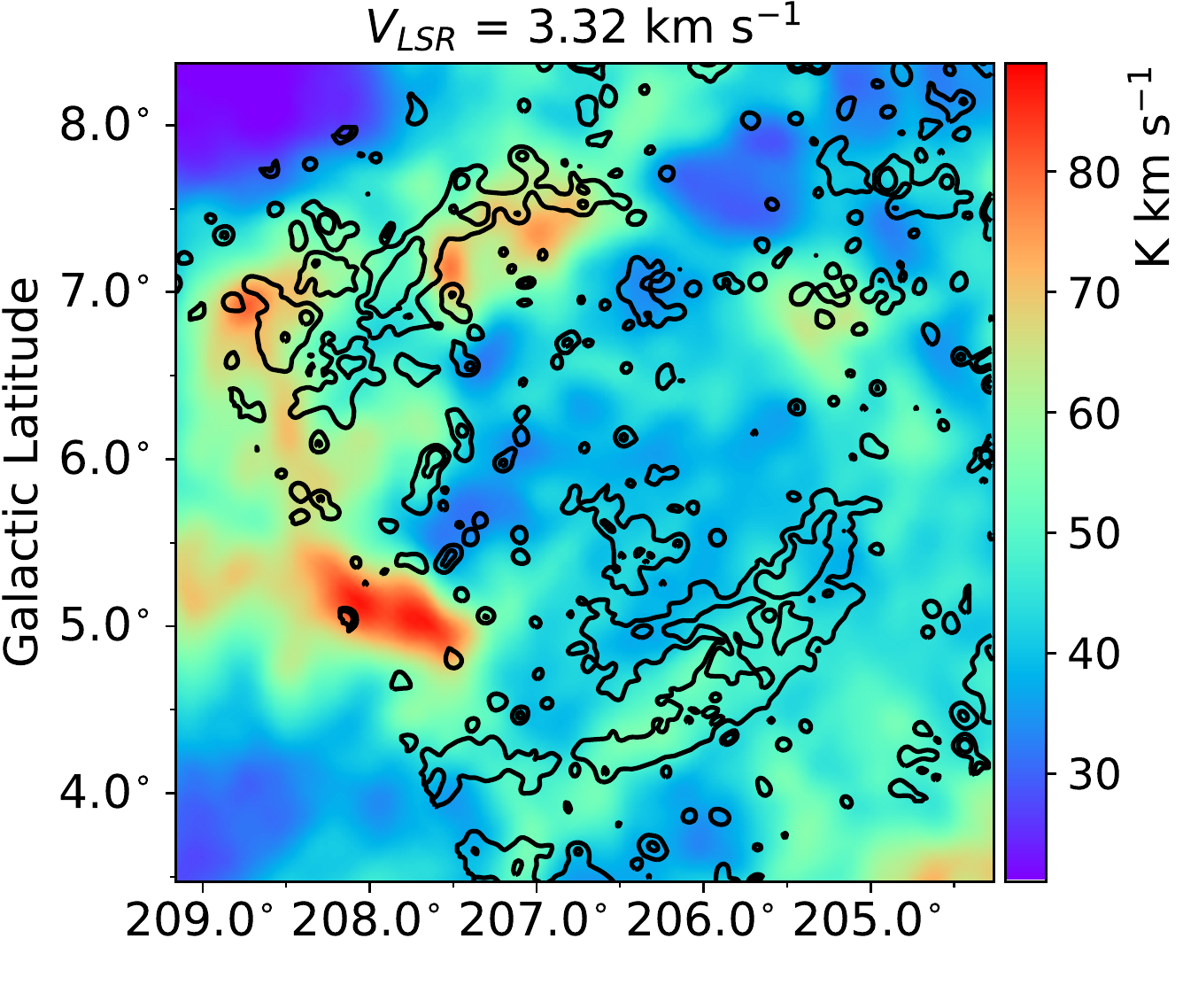}
\includegraphics[width=0.32\textwidth]{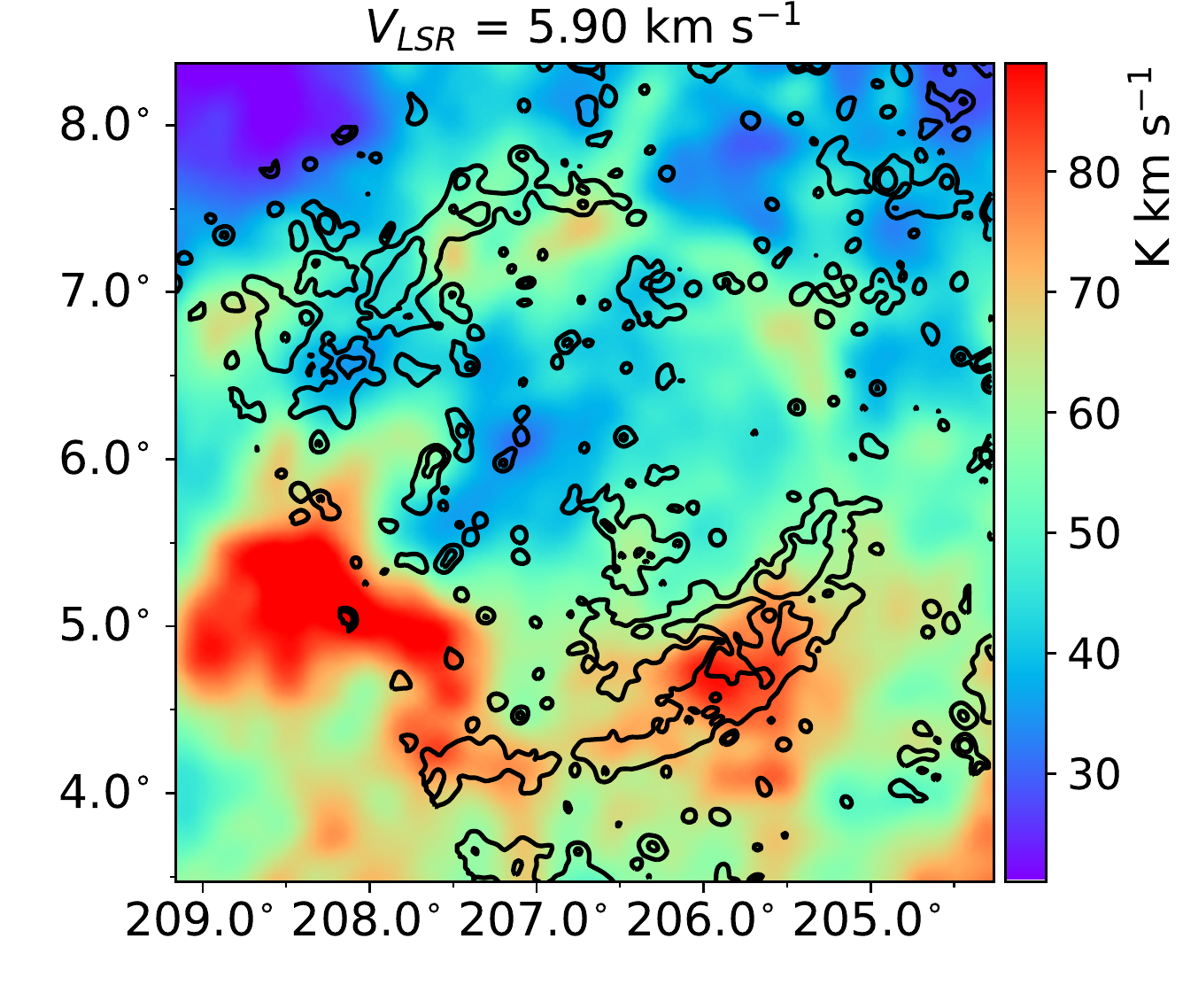}
\includegraphics[width=0.32\textwidth]{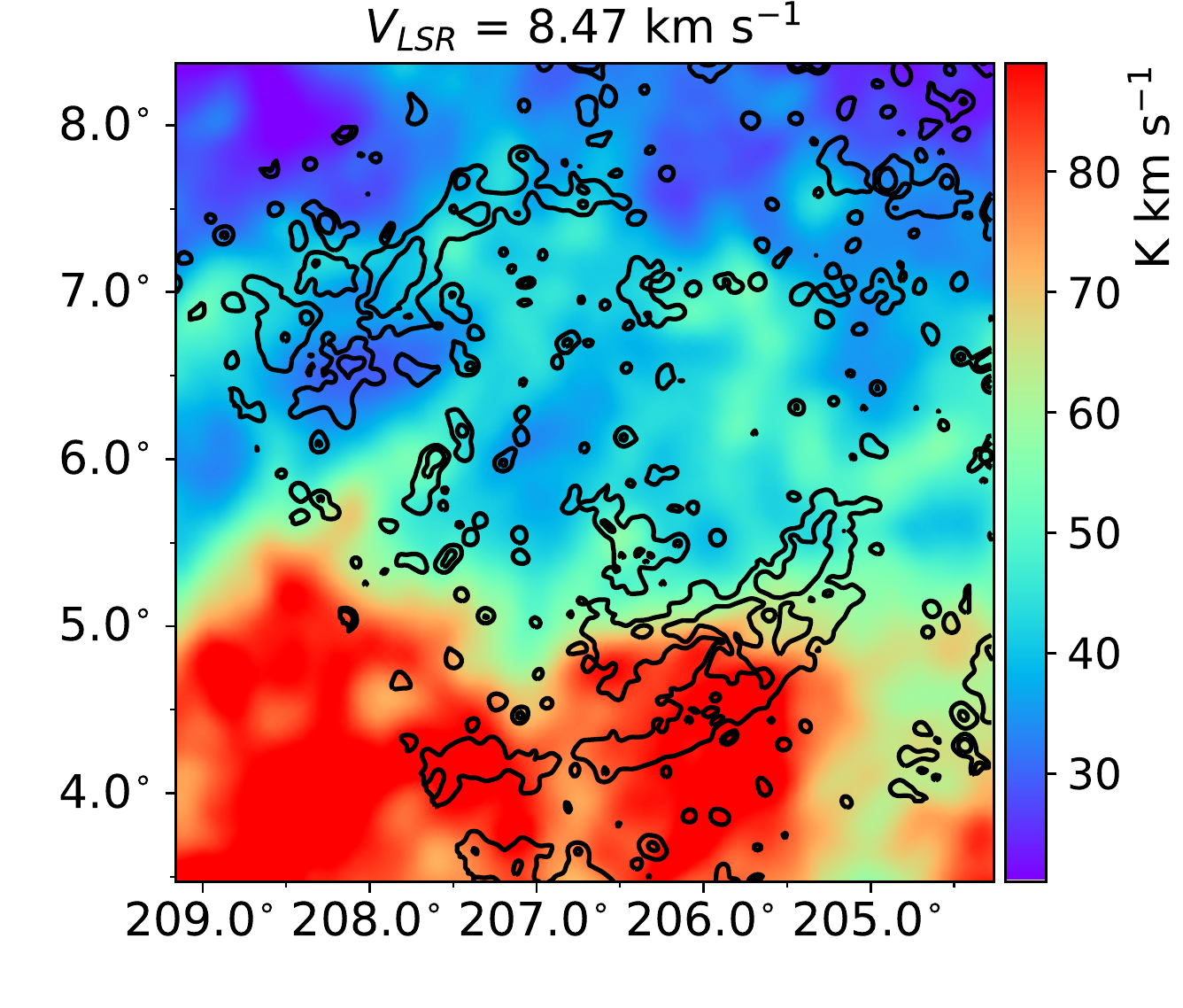}\\
\includegraphics[width=0.32\textwidth]{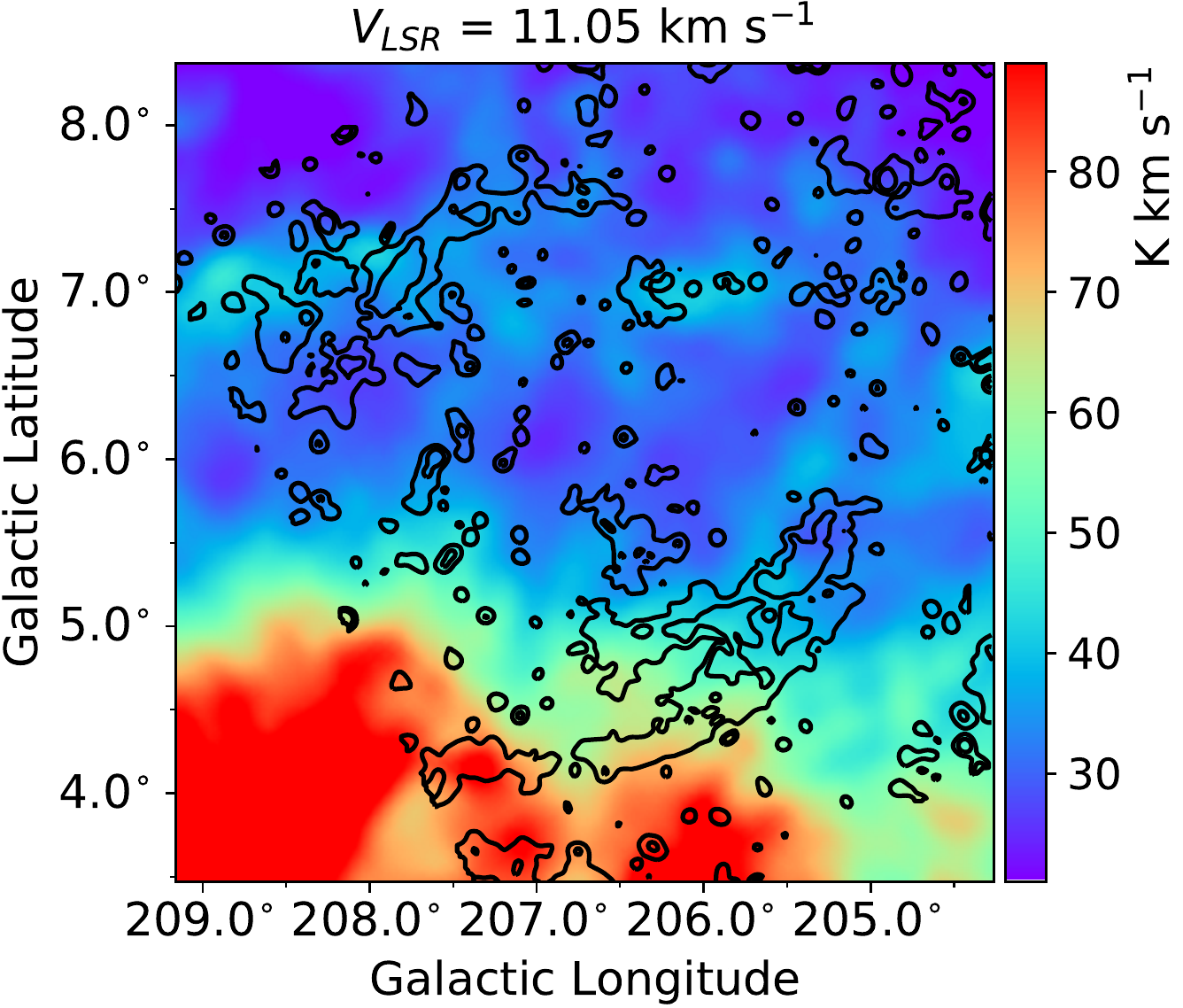}
\includegraphics[width=0.32\textwidth]{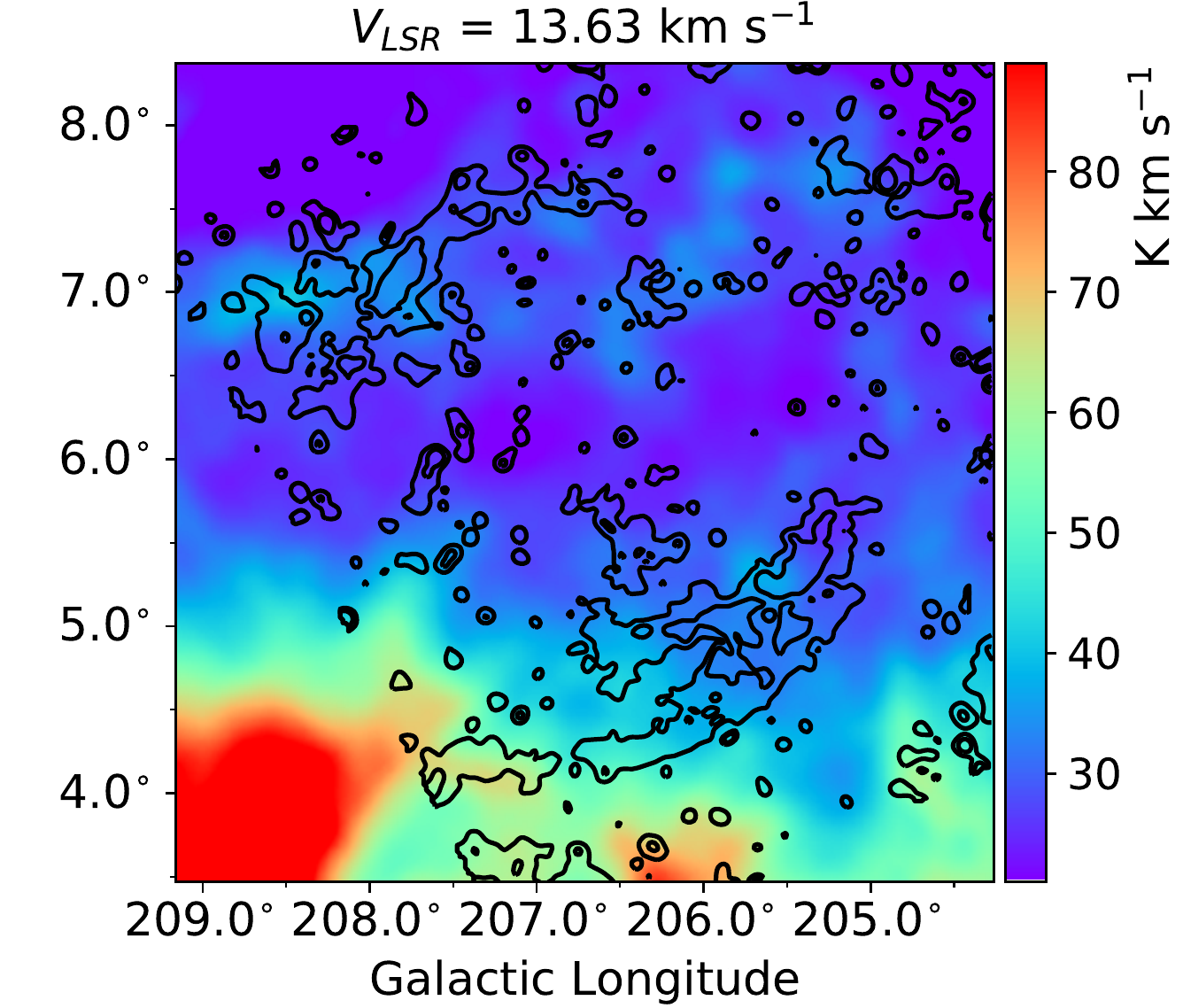}
\includegraphics[width=0.32\textwidth]{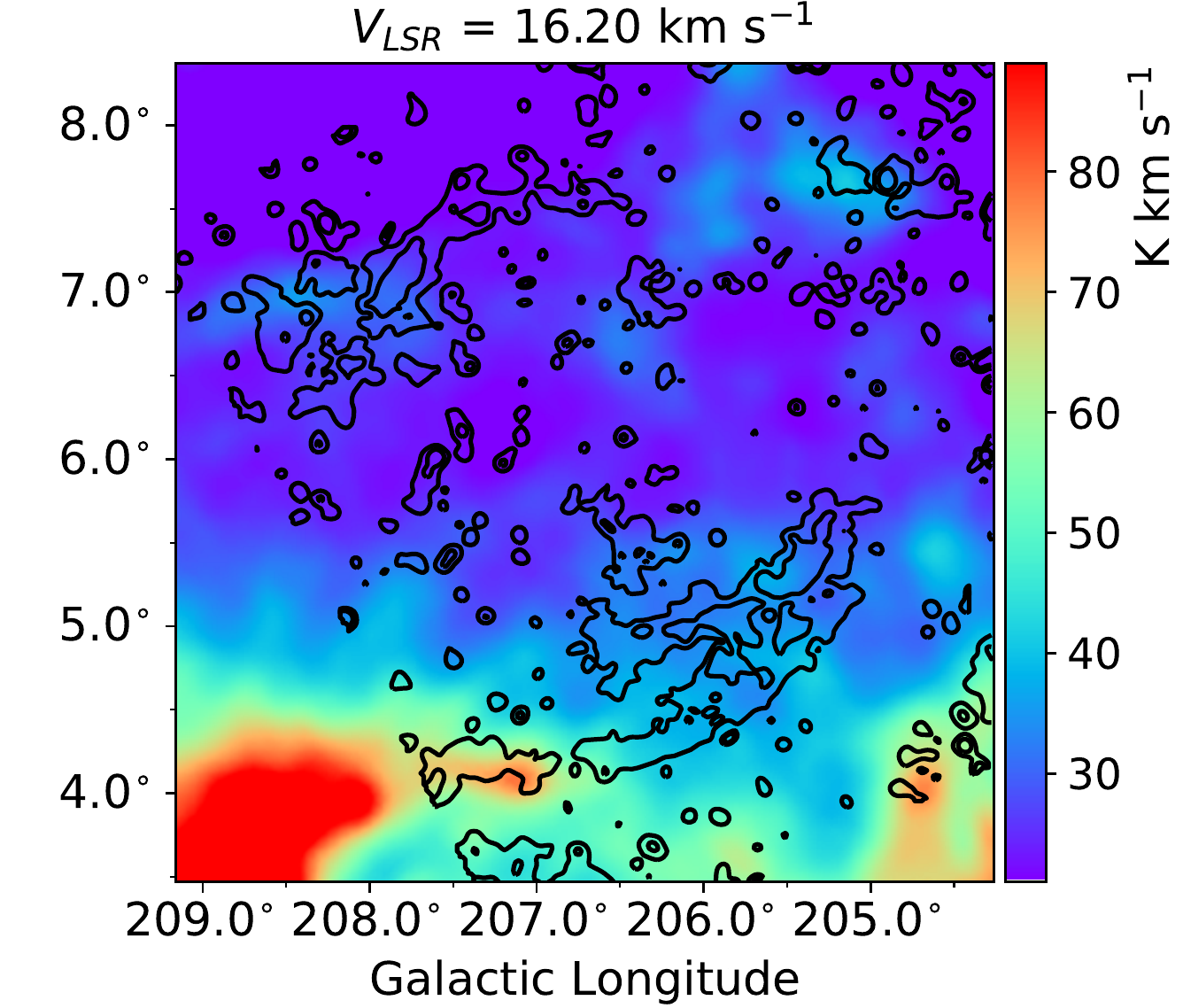}\\
\caption{Integrated \ion{H}{I} channel maps in several velocity ranges
  for the sky region of G206.7+5.9 overlaid by the FAST radio
  continuum contours. The velocity step is $\sim$2.6~km\ s$^{-1}$, and
  the central velocity is indicated on top of each panel.}
\label{fig:hi}
\end{figure*}

For comparison, we show the 4$\farcm$4-resolution Effelsberg
$\lambda$11\ cm polarization images of G203.1+6.6 and G206.7+5.9 in
the lower panel of Fig.~\ref{Fig_11PI}, at which frequency less
depolarization is expected. B-vectors as the observed magnetic-field
orientation are overlaid. Polarized emission from both G203.1+6.6 and
G206.7+5.9, although mixed with some un-related overlapping polarized
emission, can still be identified. For G203.1+6.6, two structures with
pronounced polarized emission are seen, i.e. in the center of the
image and along the arc of G203.1+6.6. A $PI$ depression at about $b =
7\degr24\arcmin$ roughly separates these two structures. It is clear
that the visible $PI$ in the center of the image is un-related to
G203.1+6.6, where no total-intensity emission was detected. For the
polarized emission in the arc region, it seems that the weak $PI$ seen
in the part of $l > 203\fdg4$ may come from the arc, while for the
area of $l < 203\fdg4$, at least part of $PI$ is not related. Because
the observed polarization intensity here exceeds the total-intensity
signal, violating the maximum percentage polarization of $\sim$75\%
for synchrotron emission. This is caused by the settings of the
relative zero-levels of the maps for $I$, and $U$, $Q$ or $PI$,
assuming zero intensities at the field edges. If the ratio $PI/I$ is
apparently above 1, the polarized emission arises at least partly
somewhere in the ISM, and is not related to the total-intensity
structure. For G206.7+5.9, intense polarized emission is perfectly
traced for the inner shell in the southwest (Galactic
coordinates). The $PI$ emission is about 13~mK $T_{B}$ at
$\lambda$11\ cm for the ``blob''-overlapped shell region. It therefore
can be extrapolated to be about 66~mK $T_{B}$ at 1.4~GHz, by using the
spectral index found in Fig.~\ref{TT-plot}. This is far less than
$\sim$180~mK $T_{B}$ detected by FAST in the same area, indicating
that at least majority of the polarized emission seen for the ``blob''
at L-band is not related to G206.7+5.9. The rotation measure of $RM$ =
29$\pm$4~rad\ m$^{-2}$ can cause a polarization-angle rotation of
75$\degr$ at L-band and heavily depolarize the polarized emission
coming from behind. If the ``blob'' is in the foreground of
G206.7+5.9, the polarized emission originated from the SNR is
therefore not likely to be seen at the L band and a clockwise rotation
of 20$\degr$ needs to be applied to the B-vectors in the
$\lambda$11\ cm image. This will better align the magnetic fields
along the inner shell of G206.7+5.9. Such parallel configuration
suggests that G206.7+5.9 is likely an evolved SNR in the adiabatic
phase \citep{Reich02}. The $\lambda$11\ cm polarized emission of the
other two shells of G206.7+5.9 is weaker, but still corresponds well
to the total-intensity emission.

In brief, the associated polarized emission traced in the shell region
of G203.1+6.6 and G206.7+5.9 at $\lambda$11\ cm adds solid proof on
the non-thermal nature of G203.1+6.6 and G206.7+5.9. A reliable
percentage polarization is not easy to obtain due to the confusion
with un-related extended Galactic polarized emission.

\subsection{Distance estimates} 

Distance is a basic parameter for astronomical objects. The distance
to SNRs can be inferred through the association with objects that have
known distances \citep[e.g. stars;][]{Humphreys1978, Fesen21} or the
kinematic method for related structures e.g. of \ion{H}{I} emission or
absorption clouds or CO emission \citep[e.g.][]{Kothes03,
  Tian07}. \citet{Kassim94} pointed out that X-ray emission can also
help to determine the distance to shell-type SNRs. In addition, dust
reddening of stars \citep[e.g.][]{Chen2017, Yu2019, Zhao2020} and red
clump stars \citep{Shan18} have been used for the distance estimates
for SNRs. We checked the ROSAT hard- (0.5-2.0~keV) and soft-band
(0.1-0.4~keV) data \citep{Voges99} toward G203.1+6.6 and
G206.7+5.9. Only a few clumps in the hard-band (0.5-2.0~keV) are
visible in the area of G206.7+5.9, but with a low signal-to-noise
ratio. We also failed to find adequate CO data currently available in
this sky area.

Morphological correlation between the radio continuum map of an SNR
and \ion{H}{I} emission structure provides another possibility in
estimating the kinematic distance of SNRs \citep[e.g.][]{Foster13,
  Kothes01}. The Effelsberg \ion{H}{I} survey \citep[hereafter
  EBHIS;][]{Winkel16} observed the sky north of $\delta \ge -5\degr$
with an angular resolution of 10$\farcm$8. The spectral resolution is
about 1.29 km\ s$^{-1}$ and the brightness temperature noise level is
about 90~mK T$_{B}$. We searched for the SNR-related structures in the
EBHIS channel maps after integrating the velocity step to
$\sim$2.6~km\ s$^{-1}$ among the entire velocity range of $-$600 to
600~km s$^{-1}$, especially in the range of $-$10 km s$^{-1}$ to
+50~km\ s$^{-1}$. For SNR G206.7+5.9, the morphological resemblance is
found between the radio continuum map and integrated \ion{H}{I} map in
the velocity range of 0-8 km s$^{-1}$, with the best correlation
seen in 3.3-5.9 km s$^{-1}$. An \ion{H}{I} cavity is noticed, which
is probably created by the progenitor star or the SN event (see in
Fig.~\ref{fig:hi}). By using the kinematic distance calculation tool
of \citet{Wenger18} based on the rotation curve and updated solar
motion parameters of \citet{Reid14}, we obtain the average value of
$V_{LSR} = 4.6~\rm km\,s^{-1}$ which corresponds to a kinematic
distance of about 0.44~kpc. This places the SNR G206.7+5.9 in the
Local Arm \citep{Reid19, Hou21} and results in a physical size of
$\sim$27\,pc.  No \ion{H}{I} structure is found to be clearly
associated with G203.1+6.6.
 
In addition, three known pulsars in the ATNF pulsar
catalog\footnote{http://www.atnf.csiro.au/research/pulsar/psrcat/}
\citep{Manchester05} are found in the field close to G203.1+6.6 and
G206.7+5.9. The pulsars J0659+1414 and J0709+0458 which are out of the
boundary of Fig.~\ref{Fig_I} have a distance estimate of
$\sim$0.29~kpc and $\sim$1.20~kpc, respectively. But if such a
scenario holds that pulsars run away from the center of the SNe,
neither of the two pulsars are related to G203.1+6.6 or G206.7+5.9,
because they cannot even be traced back to the central area of the two
SNRs based on their proper motion measurements \citep{Brisken03,
  Martinez19}. The pulsar J0711+0931 is near the eastern shell of
G206.7+5.9 and has an estimated distance of about 1.17~kpc. No proper
motion information is available yet for such discussion.

\section{Conclusions}
\label{sect:conclusion}
We conducted new scanning observations with the L-band 19-beam
receiver of FAST. Through the observations toward calibration sources,
we commissioned the FAST L-band receiving system and set up a
data-reduction pipeline which can be successfully applied to the FAST
radio-continuum observations. Two new supernova remnants G203.1+6.6
with a size of $\sim2\fdg$5 and G206.7+5.9 with a size of about
3$\fdg$5 were confirmed with the FAST L-band observations together
with the Effelsberg $\lambda$11\ cm polarization
measurements. G203.1+6.6 appears to be a single arc structure, while
G206.7+5.9 shows bilateral shells, with a double shell on one
side. The multi-channel backend of FAST enabled an in-band spectral
determination. The spectral indices of G203.1+6.6 and G206.7+5.9 were
both found to be $\beta \sim -2.6$ to $-2.7$. Significant polarized
emission was not detected toward the two SNRs at L band, but seen for
all the shell structures of G203.1+6.6 and G206.7+5.9 at
$\lambda$11\ cm. The morphology, spectral and polarization data prove
that both objects are SNRs emitting synchrotron radiation. Through
identifying morphological correlation between radio continuum and
\ion{H}{I}, we were able to make distance estimate toward G206.7+5.9.
A kinematic distance of $\sim$440~pc is determined, indicating a
physical size of about 27~pc. This distance makes G206.7+5.9 a Local
Arm object. \\

{\footnotesize {\bf Acknowledgements.} The work was supported by the
  National Key R\&D Program of China (No. 2021YFA1600401 and
  2021YFA1600400), the National Natural Science Foundation of China
  (No. 11988101), and the National SKA Program of China (Grant
  No. 2022SKA0120103). Xuyang Gao acknowledges the CAS-NWO Cooperation
  Program (Grant No. GJHZ1865), and the Open Project Program of the
  Key Laboratory of FAST, NAOC, Chinese Academy of Sciences. Xiaohui
  Sun is supported by the Cultivation Project for FAST Scientific
  Payoff and Research Achievement of CAMS-CAS, and the Science \&
  Technology. Tao Hong is supported by the National Natural Science
  Foundation of China (Grant No. 12003044). We would like to thank
  Prof. Biwei Jiang for helpful discussion on the distance of the
  SNRs. This work made use of the data from FAST (Five-hundred-meter
  Aperture Spherical radio Telescope). FAST is a Chinese national
  mega-science facility, operated by National Astronomical
  Observatories, Chinese Academy of Sciences.  This research is based
  in part on observations with the 100-m telescope of the
  Max-Planck-Institut f\"ur Radioastronomie at Effelsberg.}
%%%%%%%%%%%%%%%%%%%%%%%%%%%%%%%%%%%%%%%%%%%%%%%%%%%%%%%
%%% Acknowledgements.

%%%%%%%%%%%%%%%%%%%%%%%%%%%%%%%%%%%%%%%%%%%%%%%%%%%%%%%
\Acknowledgements{\scriptsize The work was supported by the National
  Key R\&D Program of China (No. 2021YFA1600401 and 2021YFA1600400),
  the National Natural Science Foundation of China (No. 11988101), and
  the National SKA Program of China (Grant No. 2022SKA0120103). Xuyang
  Gao acknowledges the CAS-NWO Cooperation Program (Grant
  No. GJHZ1865), and the Open Project Program of the Key Laboratory of
  FAST, NAOC, Chinese Academy of Sciences. Xiaohui Sun is supported by
  the Cultivation Project for FAST Scientific Payoff and Research
  Achievement of CAMS-CAS, and the Science \& Technology. Tao Hong is
  supported by the National Natural Science Foundation of China (Grant
  No. 12003044). We would like to thank Prof. Biwei Jiang for helpful
  discussion on the distance of the SNRs. This work made use of the
  data from FAST (Five-hundred-meter Aperture Spherical radio
  Telescope). FAST is a Chinese national mega-science facility,
  operated by National Astronomical Observatories, Chinese Academy of
  Sciences.  This research is based in part on observations with the
  100-m telescope of the Max-Planck-Institut f\"ur Radioastronomie at
  Effelsberg.}

%%%%%%%%%%%%%%%%%%%%%%%%%%%%%%%%%%%%%%%%%%%%%%%%%%%%%%%
%%% Supplements. 
%%%%%%%%%%%%%%%%%%%%%%%%%%%%%%%%%%%%%%%%%%%%%%%%%%%%%%%
%\Supplements{}
{\footnotesize \setlength{\baselineskip}{3pt}
%%%%%%%%%%%%%%%%%%%%%%%%%%%%%%%%%%%%%%%%%%%%%%%%%%%%%%%
%%%%%%%%%%%%%%%%%%%%%%%%%%%%%%%%%%%%%%%%%%%%%%%%%%%%%%%
\bibliographystyle{raa}

\bibliography{bbfile}
}
\end{multicols}

\end{document}